# On the quantum theory of light interacting with nano-materials

Masud Mansuripur

James C. Wyant College of Optical Sciences, The University of Arizona, Tucson

**Abstract**. The fundamental processes of absorption, stimulated and spontaneous emission, and elastic as well as inelastic scattering involving light and atoms, molecules, and nano-particles have been studied for decades using both classical and quantum theories. While providing an overview of the subject, this paper presents a streamlined approach to studying atom-photon interactions in the context of modern quantum optics in the hope of providing guidance for applications in the general area of molecular and nano-photonic machines.

**1. Introduction**. In the classical theory of electrodynamics, an atom, or molecule, or small particle is typically modelled as a mechanical mass-and-spring system subjected to an externally applied electromagnetic (EM) field. This so-called "Lorentz oscillator model" uses phenomenological parameters such as an effective mass, an effective charge, a spring constant, and a damping coefficient to relate the induced electric dipole moment of the particle to an applied electric field.[1-4] Maxwell's equations are subsequently solved to determine the radiated EM field produced by the oscillating dipole.[1-4] While it is certainly possible to explain many features of optical absorption, dispersion, and scattering[1-5] (even stimulated emission[6]) in the context of the classical theory, a far better understanding of these and related processes can be achieved using methods of quantum electrodynamics.[7-9] It also goes without saying that there exist natural phenomena, such as the spontaneous emission of photons by excited atoms and molecules, that are fundamentally quantum-mechanical in nature and cannot be explained without resort to the modern theory of quantum optics.[7-9] The goal of the present paper is to provide an elementary tutorial review of atom-photon interaction processes in hopes of providing guidance for applications in the general area of molecular and nano-photonic machines.

We begin in Sec.2 with a brief review of the classical Hamiltonian for a point-charge in the presence of an external EM field. Section 3 introduces a quantum-mechanical version of this Hamiltonian and proceeds to break it up into a particle Hamiltonian plus an interaction Hamiltonian (between the particle and the quantized external EM radiation). Whereas the particle Hamiltonian is responsible for creating the electronic eigenstates of the atom under consideration, while the radiation Hamiltonian defines the energy eigenstates of the various modes of the free EM field, it is the interaction Hamiltonian that governs the co-evolution of the states of the atom and the radiation field. In preparation for an analysis of atom-photon interactions, Sec.4 introduces the notion of the density of states (i.e., the number of field modes per unit energy per unit solid angle) for the propagating EM radiation in free space. The modes of the free EM field are hereinafter identified as $(\omega, \boldsymbol{k}, \hat{\boldsymbol{e}})$, where $\omega$ is the temporal frequency, $\boldsymbol{k} = (\omega/c)\hat{\boldsymbol{\kappa}}$ is the wave-vector, $c$ is the speed of light in vacuum, and the (generally complex-valued) unit-vector $\hat{\boldsymbol{e}}$ specifies the polarization of the mode. For each frequency $\omega$ and (real-valued) unit-vector $\hat{\boldsymbol{\kappa}}$, there exist two independent modes, one with polarization $\hat{\boldsymbol{e}}_1$, the other with an orthogonal polarization $\hat{\boldsymbol{e}}_2$, with $\hat{\boldsymbol{\kappa}} \cdot \hat{\boldsymbol{e}}_1 = \hat{\boldsymbol{\kappa}} \cdot \hat{\boldsymbol{e}}_2 = 0$ and $\hat{\boldsymbol{e}}_1 \cdot \hat{\boldsymbol{e}}_2^* = 0$. A useful identity involving the triplet of unit-vectors $(\hat{\boldsymbol{\kappa}}, \hat{\boldsymbol{e}}_1, \hat{\boldsymbol{e}}_2)$ is presented in the Appendix.

Elastic scattering via atom-photon interaction in accordance with first-order perturbation theory is the subject of Sec.5. Here, an incident photon of frequency $\omega$ and wave-vector $\hat{\boldsymbol{k}}_1 = (\omega/c)\hat{\boldsymbol{\kappa}}_1$ is scattered into a different direction $\hat{\boldsymbol{\kappa}}_2$ while retaining its energy $\hbar\omega$. This happens without a change in the state $|\psi\rangle$ of the scattering particle, hence the designation "elastic scattering". Section 6 is devoted to an analysis of the processes of absorption, stimulated emission, and spontaneous emission. Absorption occurs when an atom captures a photon from the incident radiation and transitions to a higher-energy state. In the case of stimulated emission, an atom in an exited state releases a photon into an incident radiation mode $(\omega, \boldsymbol{k}, \hat{\boldsymbol{e}})$, raising the number of photons in the mode by 1, while the atom transitions to a correspondingly lower-energy state. Spontaneous emission is similar to stimulated emission except that the emitted photon enters an empty mode of the EM field—i.e., a vacuum mode, which is devoid of photons initially. The stimulation of the excited atom in this case is done by vacuum fluctuations.[7]

In Sec.7, we examine elastic as well as inelastic scattering of photons by atoms in accordance with second-order perturbation theory; the cases of Rayleigh, Thomson, and resonant scattering are treated in this section. An alternative (and widely-used) interaction Hamiltonian, known as the electric-dipole Hamiltonian, is introduced in Sec.8, then used in Sec.9 to relate the atomic absorption and emission rates to the corresponding electric-dipole interaction coefficient. In Sec.10, we compute the cross-section of an atom in cases of resonant scattering. Multiphoton processes are addressed in Sec.11, where the nonlinear optical phenomena of 2-photon absorption and stimulated emission, and also the mechanism of 2-photon spontaneous emission are discussed. In Sec.12, we describe polarization-entangled photon pairs emitted in an atomic radiative cascade. Section 13 is devoted to a discussion of atomic energy-level-shifts due to interactions of the atom with vacuum fluctuations. The paper closes with a brief summary and a few concluding remarks in Sec.14.

**2. Classical Hamiltonian.** Consider a point-particle of mass $m$ and charge $q$ moving in space-time $(\boldsymbol{r}, t)$ under the influence of the scalar potential $\Phi(\boldsymbol{r}, t)$ and the vector potential $\boldsymbol{A}(\boldsymbol{r}, t)$. Denoting by $\boldsymbol{p} = p_x \hat{\boldsymbol{x}} + p_y \hat{\boldsymbol{y}} + p_z \hat{\boldsymbol{z}}$ the *canonical* momentum of the particle, the Hamiltonian of the system may be written as follows:

$$H(\boldsymbol{r}, \boldsymbol{p}, t) = \frac{[\boldsymbol{p} - q\boldsymbol{A}(\boldsymbol{r}, t)]^2}{2m} + q\Phi(\boldsymbol{r}, t). \tag{1}$$

The equations of motion are now derived from the Hamiltonian following standard procedure:

$$\dot{x} = \frac{\partial H}{\partial p_x} = \frac{2(\partial \boldsymbol{p}/\partial p_x) \cdot [\boldsymbol{p} - q\boldsymbol{A}(\boldsymbol{r}, t)]}{2m} = \frac{p_x - qA_x(\boldsymbol{r}, t)}{m}, \tag{2a}$$

$$\dot{p}_x = -\frac{\partial H}{\partial x} = \frac{q[\boldsymbol{p} - q\boldsymbol{A}(\boldsymbol{r}, t)] \cdot \partial \boldsymbol{A}(\boldsymbol{r}, t)/\partial x}{m} - q\frac{\partial \Phi(\boldsymbol{r}, t)}{\partial x}. \tag{2b}$$

Similar equations, of course, hold for the $y$ and $z$ coordinates. From Eq.(2a) we now find

$$p_x = m\dot{x} + qA_x(\boldsymbol{r}, t) \quad \rightarrow \quad \dot{p}_x = m\ddot{x} + q\left[\left(\frac{\partial A_x}{\partial x}\right)\dot{x} + \left(\frac{\partial A_x}{\partial y}\right)\dot{y} + \left(\frac{\partial A_x}{\partial z}\right)\dot{z} + \frac{\partial A_x}{\partial t}\right]. \tag{3}$$

Substitution into Eq.(2b) yields

$$m\ddot{x} + q\left[\left(\frac{\partial A_x}{\partial x}\right)\dot{x} + \left(\frac{\partial A_x}{\partial y}\right)\dot{y} + \left(\frac{\partial A_x}{\partial z}\right)\dot{z} + \frac{\partial A_x}{\partial t}\right] = q\frac{\partial \boldsymbol{A}}{\partial x} \cdot \dot{\boldsymbol{r}} - q\frac{\partial \Phi}{\partial x}$$

$$\rightarrow \quad m\ddot{x} = q\left(-\frac{\partial \Phi}{\partial x} - \frac{\partial A_x}{\partial t}\right) + q\dot{y}\left(\frac{\partial A_y}{\partial x} - \frac{\partial A_x}{\partial y}\right) + q\dot{z}\left(\frac{\partial A_z}{\partial x} - \frac{\partial A_x}{\partial z}\right)$$

$$\rightarrow \quad m\ddot{x} = q(E_x + \dot{y}B_z - \dot{z}B_y) \quad \rightarrow \quad m\ddot{\boldsymbol{r}} = q(\boldsymbol{E} + \dot{\boldsymbol{r}} \times \boldsymbol{B}). \tag{4}$$

The last equation, of course, is Newton's equation of motion for a particle of mass $m$ and charge $q$ under the influence of the Lorentz force exerted by the $E$- and $B$-fields of an external EM field. Note that the particle motion is non-relativistic, the gauge is arbitrary, and the canonical momentum $\boldsymbol{p}$ differs from the kinetic momentum $m\dot{\boldsymbol{r}}$. The Hamiltonian of Eq.(1) is the sum of the kinetic and potential energies of the particle, but it excludes the EM field energy. If this Hamiltonian is used in a quantum-mechanical analysis of the system, the particle will be treated quantum-mechanically, but the field will remain classical, resulting in a quasi-classical treatment.[8]

**3. Quantum Hamiltonian**. A quantum treatment of the EM field requires the addition of the field Hamiltonian, namely, $\hat{\mathcal{H}}_{\text{field}} = \sum_j \hbar\omega_j(\hat{a}_j^\dagger \hat{a}_j + \tfrac{1}{2})$, to that of Eq.(1).[7-9] Here, the sum is over all the modes $j$ of the EM field. In free space, the field modes are typically propagating plane-waves that occupy a large spatial volume $V = L^3$ within the $xyz$-space, having discretized $k$-vectors $\boldsymbol{k} = (2\pi/L)(n_x \hat{\boldsymbol{x}} + n_y \hat{\boldsymbol{y}} + n_z \hat{\boldsymbol{z}})$, temporal frequencies $\omega = c|\boldsymbol{k}|$, and polarization states identified by one or the other of the complex unit-vectors $\hat{\boldsymbol{e}}_1$ and $\hat{\boldsymbol{e}}_2$, as depicted in Fig.1. As usual, $n_x, n_y, n_z$ are arbitrary integers (positive, zero, or negative), $\omega$ is non-negative, $\boldsymbol{k} \cdot \hat{\boldsymbol{e}}_1 = \boldsymbol{k} \cdot \hat{\boldsymbol{e}}_2 = 0$, and $\hat{\boldsymbol{e}}_1 \cdot \hat{\boldsymbol{e}}_1^* = \hat{\boldsymbol{e}}_2 \cdot \hat{\boldsymbol{e}}_2^* = 1$, while $\hat{\boldsymbol{e}}_1 \cdot \hat{\boldsymbol{e}}_2^* = 0$.[†]

Suppose the immobile nucleus of an atom centered at the origin of coordinates has positive charge $Ze$, and that a single electron of mass $m$, charge $q = -e$, and position $\boldsymbol{r}(t)$ is bound to the nucleus via the Coulomb potential $\Phi(\boldsymbol{r}) = Ze/(4\pi\varepsilon_0 r)$ of the nucleus. In the absence of an external EM field, the single electron's Hamiltonian will be $\hat{\mathcal{H}}_{\text{particle}} = \hat{\boldsymbol{p}}^2/(2m) - Ze^2/(4\pi\varepsilon_0 \hat{r})$, where $\hat{\boldsymbol{p}}$, the canonical momentum operator, equals $-i\hbar\boldsymbol{\nabla}$ in position-space representation, while $\hat{r}$ is the position operator corresponding to the electron's distance from the nucleus. The energy eigenstates of this hydrogen-like atom are solutions of the Schrödinger equation $i\hbar\partial_t|\psi\rangle = \hat{\mathcal{H}}_{\text{particle}}|\psi\rangle$. The remaining terms of the Hamiltonian of Eq.(1) now form the interaction Hamiltonian between the atom and the external EM field,[9] as follows:

$$\hat{\mathcal{H}}_{\text{interaction}} = (e/m)\hat{\boldsymbol{p}} \cdot \hat{\boldsymbol{A}} + (e^2/2m)\hat{\boldsymbol{A}} \cdot \hat{\boldsymbol{A}}. \tag{5}$$

In arriving at the above equation, we have invoked the commutativity of $\hat{\boldsymbol{p}}$ and $\hat{\boldsymbol{A}}(\boldsymbol{r})$, rooted in the fact that $(\boldsymbol{\nabla} e^{i\boldsymbol{k}\cdot\boldsymbol{r}}) \cdot \hat{\boldsymbol{e}}_{1,2} = (i\boldsymbol{k} \cdot \hat{\boldsymbol{e}}_{1,2})e^{i\boldsymbol{k}\cdot\boldsymbol{r}} = 0$. Assuming the radiation wavelength is substantially greater than the atomic radius, one may substitute the vector potential operator at the origin of coordinates (i.e., at the seat of the nucleus, where $\boldsymbol{r} = 0$) for the one acting on the atom's electron located at $\boldsymbol{r}(t)$; that is, $\hat{\boldsymbol{A}} \cong \sum_j \sqrt{\hbar/(2\varepsilon_0 V \omega_j)}\,(\hat{\boldsymbol{e}}_j \hat{a}_j + \hat{\boldsymbol{e}}_j^* \hat{a}_j^\dagger)$. This is generally referred to as the long-wavelength approximation.[7-9]

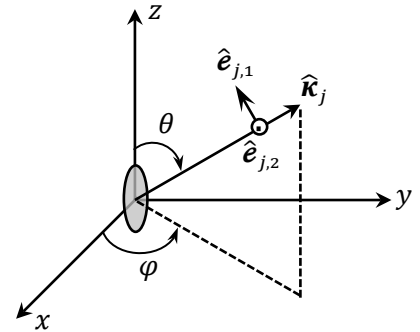

**Fig.1**. Sitting at the origin of the coordinate system is a hydrogen-like atom that interacts with the EM radiation in mode $j$, which has frequency $\omega_j$, propagates along the $k$-vector $\boldsymbol{k}_j = (\omega_j/c)\hat{\boldsymbol{\kappa}}_j$, and could be polarized along either of the unit-vectors $\hat{\boldsymbol{e}}_{j,1}$ and $\hat{\boldsymbol{e}}_{j,2}$. The polar and azimuthal angles of $\boldsymbol{k}_j$ are identified as $\theta$ and $\varphi$, respectively. The expected position vector $\boldsymbol{r}_{12}$ of the electron during a transition between the atomic energy eigenstates $|\psi\rangle$ and $|\chi\rangle$, namely, $\boldsymbol{r}_{12} = \langle\hat{r}\rangle_{\psi\to\chi} = \langle\chi|\hat{r}|\psi\rangle$, may be assumed to be along the $z$-axis. For a linearly-polarized radiation mode, where $\hat{\boldsymbol{e}}_1$ and $\hat{\boldsymbol{e}}_2$ are real-valued, one may take $\hat{\boldsymbol{e}}_1$ to be coplanar with $\boldsymbol{k}$ and $\boldsymbol{r}_{12}$, in which case $\hat{\boldsymbol{e}}_2$ will be perpendicular to the plane that contains $\hat{\boldsymbol{e}}_1, \boldsymbol{k}$, and $\boldsymbol{r}_{12}$.

**Warning**: The conventional use of carets over unit-vectors such as $\hat{\boldsymbol{e}}_j$, and also over operators such as $\hat{a}_j$, could be confusing. One should be able to tell them apart by paying close attention to the context in which these caret-adorned symbols appear. Another potential source of confusion is our use of the letter $e$ for the charge of an electron and also for the unit-vector $\hat{\boldsymbol{e}}$ of the EM field's polarization. There should be no ambiguity, however, since the symbol used for the unit-vector is bold-faced and adorned with a caret.

---

[†] The mutually orthogonal polarization vectors $\hat{\boldsymbol{e}}_1$ and $\hat{\boldsymbol{e}}_2$ automatically satisfy the condition $\boldsymbol{k} \times (\hat{\boldsymbol{e}}_1 \times \hat{\boldsymbol{e}}_2) = 0$, since this triple product equals $(\boldsymbol{k} \cdot \hat{\boldsymbol{e}}_2)\hat{\boldsymbol{e}}_1 - (\boldsymbol{k} \cdot \hat{\boldsymbol{e}}_1)\hat{\boldsymbol{e}}_2 = 0$. Also, $\hat{\boldsymbol{e}}_1 \times \hat{\boldsymbol{e}}_2$ is a unit-vector—albeit with a complex magnitude, in general—since $(\hat{\boldsymbol{e}}_1 \times \hat{\boldsymbol{e}}_2) \cdot (\hat{\boldsymbol{e}}_1^* \times \hat{\boldsymbol{e}}_2^*) = (\hat{\boldsymbol{e}}_1 \cdot \hat{\boldsymbol{e}}_1^*)(\hat{\boldsymbol{e}}_2 \cdot \hat{\boldsymbol{e}}_2^*) - (\hat{\boldsymbol{e}}_1 \cdot \hat{\boldsymbol{e}}_2^*)(\hat{\boldsymbol{e}}_2 \cdot \hat{\boldsymbol{e}}_1^*) = 1$. Aside from an inconsequential phase-factor, $\hat{\boldsymbol{e}}_1 \times \hat{\boldsymbol{e}}_2$ is thus seen to be a unit-vector aligned with $\hat{\boldsymbol{\kappa}} = c\boldsymbol{k}/\omega$.

**Digression**: In practice, the electric-dipole Hamiltonian $\widehat{\mathcal{H}}_{\text{interaction}} = e\hat{\boldsymbol{r}} \cdot \widehat{\boldsymbol{E}}$ is frequently used in place of the interaction Hamiltonian of Eq.(5); here, $\hat{\boldsymbol{r}}$ is the position operator for the electron (charge $= -e$) and, in the long-wavelength approximation, $\widehat{\boldsymbol{E}} \cong \sum_j \mathrm{i}\sqrt{\hbar\omega_j/(2\varepsilon_0 V)}\,(\hat{\boldsymbol{e}}_j\hat{a}_j - \hat{\boldsymbol{e}}_j^*\hat{a}_j^\dagger)$ is the (transverse) $E$-field operator at the origin of the coordinates (see Sec.8). Note that both operators $\widehat{\boldsymbol{A}}$ and $\widehat{\boldsymbol{E}}$ used in the interaction Hamiltonian are in the Schrödinger picture; that is, the time-dependence factors $e^{\pm \mathrm{i}\omega t}$ that accompany these operators in the Heisenberg picture are left out.[9] Later, when we invoke time-dependent perturbation theory to compute the transition probabilities associated with the processes of photon absorption, emission, and scattering by the atom, we shall work in the Schrödinger picture by calling upon the time-dependent states of the atom and the radiation field.

**4. Density of radiation states**. For any given $\boldsymbol{k}$ occupying a $k$-space volume element $(2\pi/L)^3$, the quasi-continuum of the EM field in free space consists of two states, one for each of the polarizations $\hat{\boldsymbol{e}}_1$ and $\hat{\boldsymbol{e}}_2$. Working in the spherical coordinate system $(k, \theta, \varphi)$, the number of discrete $k$-vectors inside the differential volume $\mathrm{d}k\mathrm{d}\Omega$ is found to be $(L/2\pi)^3 k^2 \mathrm{d}k\mathrm{d}\Omega$. Here $\mathrm{d}\Omega = \sin\theta\,\mathrm{d}\theta\mathrm{d}\varphi$ is the solid angle subtended by the differential volume at the origin of the $k$-space. Considering that a single photon's energy in the $(\omega_\ell, \boldsymbol{k}_\ell, \hat{\boldsymbol{e}}_\ell)$ mode is $\mathcal{E} = \hbar\omega_\ell = \hbar c k_\ell$, the number of discrete modes having energies in the differential interval $(\mathcal{E}, \mathcal{E} + \mathrm{d}\mathcal{E})$ in the immediate neighborhood of $(\omega_\ell, \boldsymbol{k}_\ell, \hat{\boldsymbol{e}}_\ell)$ within the $k$-space turns out to be $(L/2\pi\hbar c)^3 \mathcal{E}^2 \mathrm{d}\mathcal{E}\mathrm{d}\Omega$. The corresponding density of states (per unit energy per unit solid angle) is thus seen to be

$$\rho_s(\theta, \varphi; \mathcal{E}) = V\mathcal{E}^2/(2\pi\hbar c)^3. \tag{6}$$

This expression of the density of radiation states in vacuum is obtained under the assumption that the vacuum modes are unconstrained by any engineered structures (e.g., waveguides, cavities), and that, for each mode, only one polarization ($\hat{\boldsymbol{e}}_1$ or $\hat{\boldsymbol{e}}_2$) is allowed. The isotropy of free space accounts for the fact that $\rho_s$ does *not* depend on the propagation direction $(\theta, \varphi)$ of the considered modes. Finally, the quadratic dependence of $\rho_s$ on photon energies ($\mathcal{E} = \hbar\omega$) is a consequence of the quadratic increase with distance from the origin of the volume element of the $k$-space that subtends the differential solid angle $\mathrm{d}\Omega$ at the origin of coordinates.

**5. Elastic scattering caused by atom-photon interaction in first-order perturbation theory**. The operator $(e^2/2m)\widehat{\boldsymbol{A}} \cdot \widehat{\boldsymbol{A}}$ in Eq.(5) contains terms such as $[e^2\hbar/(2\varepsilon_0 mV\omega_j)](\hat{a}_j^\dagger \hat{a}_j + \tfrac{1}{2})$, which give rise to diagonal matrix elements and should, therefore, be added to the radiation Hamiltonian $\widehat{\mathcal{H}}_{\text{field}}$. However, considering that these terms are far smaller than the corresponding terms $\hbar\omega_j(\hat{a}_j^\dagger \hat{a}_j + \tfrac{1}{2})$ of the radiation Hamiltonian, they may be safely ignored. Of the remaining terms, the operator

$$[e^2\hbar/(4\varepsilon_0 mV\sqrt{\omega_j\omega_\ell})][(\hat{\boldsymbol{e}}_j \cdot \hat{\boldsymbol{e}}_\ell^*)\hat{a}_j\hat{a}_\ell^\dagger + (\hat{\boldsymbol{e}}_\ell^* \cdot \hat{\boldsymbol{e}}_j)\hat{a}_\ell^\dagger\hat{a}_j] \tag{7}$$

contributes to an elastic scattering process whereby a photon from mode $j$ is scattered into mode $\ell$ — without altering the mediating atom's state. Conservation of energy then demands that the incident and scattered photons possess the same energy; that is, $\hbar\omega_j = \hbar\omega_\ell$. Given that the first term $(e/m)\widehat{\boldsymbol{p}} \cdot \widehat{\boldsymbol{A}}$ of the interaction Hamiltonian of Eq.(5) cannot contribute to photon scattering in first-order perturbation theory, the operator appearing in Eq.(7) is the only one that can bring about the annihilation of a photon in mode $j$ while creating a photon in mode $\ell$ *without* resort to higher-order perturbations. Thus, staying with first-order perturbation theory for the time being, we denote the state of the atom (both before and after the scattering event) by $|\psi\rangle$, and the initial number of photons in modes $j$ and $\ell$ by $n_j$ and $n_\ell$, respectively. The operator in Eq.(7) then yields the matrix element for transition from $|\psi; n_j, n_\ell\rangle$ to $|\psi; n_j - 1, n_\ell + 1\rangle$, as follows:

$$w = e^2\hbar(\hat{\boldsymbol{e}}_j \cdot \hat{\boldsymbol{e}}_\ell^*)\sqrt{n_j(n_\ell + 1)}/(2\varepsilon_0 mV\omega_j). \tag{8}$$

The scattering probability is thus predicted to rise with $n_j$, which is proportional to the intensity of the EM field in mode $j$. Also, when $n_\ell \neq 0$, the scattering probability into mode $\ell$ is predicted to rise with the product of the intensities of the $j$ and $\ell$ modes — this being a form of stimulated and nonlinear scattering.

In cases where an incident photon scatters into an initially empty mode $\ell$ (i.e., $n_\ell = 0$), assuming the accessible vacuum modes surrounding mode $\ell$ are not deliberately suppressed or constrained, transitions from the initial state $|\psi; n_j, n_\ell = 0\rangle$ to the quasi-continuum surrounding the mode $\ell$ will be governed by Fermi's golden rule,[8,9] where the transition rate is given by

$$\Gamma = 2\pi |w|^2 \rho_s(\theta, \varphi; \mathcal{E})/\hbar. \tag{9}$$

Setting $n_\ell = 0$ in the expression of $w$ in Eq.(8), recalling that energy conservation dictates the equality of the incident and scattered photon energies in this elastic scattering process (i.e., $\mathcal{E} = \hbar\omega_\ell = \hbar\omega_j$), and using Eq.(6) for the density of states $\rho_s$ per unit energy per unit solid angle (for the only scattered photon polarization $\hat{e}_\ell$ that is presently under consideration), we arrive at

$$\frac{d\Gamma}{d\Omega} = \left(\frac{n_j c}{V}\right)\left(\frac{e^2}{4\pi\varepsilon_0 mc^2}\right)^2 |\hat{e}_j \cdot \hat{e}_\ell^*|^2. \tag{10}$$

Here, $d\Gamma/d\Omega$ is the probability per unit time per unit solid angle that a single photon of polarization $\hat{e}_j$ from the incident state $|n_j\rangle$ scatters into empty modes surrounding $\hat{\kappa}_\ell = (\theta, \varphi)$ with polarization $\hat{e}_\ell$. The leading coefficient $n_j c/V$ in Eq.(10) signifies the rate at which the $n_j$ photons of mode $j$ cross a unit area per unit time. This is because the volume taken up by a single photon in one second while crossing a unit area at the speed of light $c$ is a fraction $c/V$ of the entire volume $V$ occupied by the photons of mode $j$. Finally, the factor $e^2/(4\pi\varepsilon_0 mc^2)$ is the classical electron diameter $2r_e \cong 2.8$ fm.[‡] The differential cross-sectional area per unit solid angle for each incident photon in this kind of scattering is seen to be $d\sigma/d\Omega = (2r_e)^2 |\hat{e}_j \cdot \hat{e}_\ell^*|^2$.

In the special case when $\hat{\kappa}_j = \hat{y}$, $\hat{e}_j = \hat{z}$, and $\hat{\kappa}_\ell$ is in the $(\theta, \varphi)$ direction, if $\hat{e}_{\ell,1}$ is taken to be in the plane of $\hat{\kappa}_\ell$ and $\hat{z}$, we will have $|\hat{e}_j \cdot \hat{e}_{\ell,1}^*| = \sin\theta$. The other allowed polarization $\hat{e}_{\ell,2}$ of the scattered photon will then be perpendicular to the plane of $\hat{\kappa}_\ell$ and $\hat{z}$, resulting in $\hat{e}_j \cdot \hat{e}_{\ell,2}^* = 0$. Under such circumstances, the overall scattering cross-section for each incident photon in the absence of contributions by second-order (and higher) perturbations will be

$$\sigma = \iint_{\text{all space}} (2r_e)^2 |\hat{e}_j \cdot \hat{e}_{\ell,1}^*|^2 d\Omega = (2r_e)^2 \int_{\varphi=0}^{2\pi} \int_{\theta=0}^{\pi} \sin^3\theta \, d\theta d\varphi = (8\pi/3)(2r_e)^2. \tag{11}$$

We will see further below that this type of elastic scattering dominated by the $(e^2/2m)\hat{A} \cdot \hat{A}$ interaction Hamiltonian of Eq.(5) is characteristic of the so-called Thomson scattering, which occurs when the incident photon energy is well above the ionization energy of the atom (i.e., falls in the deep UV and soft $X$ ray regimes).

**6. Quantum-mechanical processes of photon absorption and photon emission by atoms**. The first term in the interaction Hamiltonian of Eq.(5) is responsible for a number of physical phenomena of fundamental importance including absorption, spontaneous and stimulated emission, and elastic as well as inelastic scattering. The momentum operator $\hat{p}$ acts on the atomic electron's wavefunction, say, $\psi(r)$, to produce the function $-i\hbar\nabla\psi(r)$, whose parity (odd or even) is opposite that of $\psi(\boldsymbol{r})$. Upon dot-multiplication into the polarization vector ($\hat{\boldsymbol{e}}_j$ or $\hat{\boldsymbol{e}}_j^*$) of a radiation mode $j$, the vectorial function $\boldsymbol{\nabla}\psi$ becomes the scalar function $(\boldsymbol{\nabla}\psi)\cdot\hat{\boldsymbol{e}}_j$ or $(\boldsymbol{\nabla}\psi)\cdot\hat{\boldsymbol{e}}_j^*$, which must possess certain symmetries in the $xyz$ space if this particular radiation mode is to have any $\hat{\boldsymbol{p}}\cdot\hat{\boldsymbol{A}}$-related interaction at all with the atom in the state $|\psi\rangle$. It is clear that the diagonal matrix element $\langle\psi|\hat{\boldsymbol{p}}\cdot\hat{\boldsymbol{e}}_j|\psi\rangle$ or $\langle\psi|\hat{\boldsymbol{p}}\cdot\hat{\boldsymbol{e}}_j^*|\psi\rangle$ vanishes in any case, simply because $\psi(\boldsymbol{r})$ and its gradient $\boldsymbol{\nabla}\psi(\boldsymbol{r})$ have opposite parities. As for the off-diagonal matrix elements, they will be nonzero provided that the atomic state $|\psi\rangle$ transitions to a different state $|\chi\rangle$ of opposite parity and that $\langle\chi|\hat{\boldsymbol{p}}\cdot\hat{\boldsymbol{e}}_j|\psi\rangle \neq 0$ or $\langle\chi|\hat{\boldsymbol{p}}\cdot\hat{\boldsymbol{e}}_j^*|\psi\rangle \neq 0$.

---

[‡] The $E$-field produced by a uniformly-charged hollow spherical shell of radius $r_e$ and total charge $-e$ at a distance $r$ from the center of the sphere is $E(r) = -e\hat{r}/(4\pi\varepsilon_0 r^2)$. Integrating the $E$-field energy density $\frac{1}{2}\varepsilon_0 E^2$ from $r = r_e$ to $\infty$ yields $e^2/(8\pi\varepsilon_0 r_e)$, which, upon equating with the mass energy $mc^2$ of an electron, yields $r_e = e^2/(8\pi\varepsilon_0 mc^2)$.

The vector potential operator $\hat{\boldsymbol{A}}$ consists of a sum over the annihilation and creation operators $\hat{a}_j$ and $\hat{a}_j^\dagger$ of all the modes $j$ of the EM radiation. Consequently, in first-order perturbation theory, the action of $\hat{\boldsymbol{p}}\cdot\hat{\boldsymbol{A}}$ on the joint state $|\psi; n_j\rangle$ of an atom in the state $|\psi\rangle$ plus a radiation mode $j$ in the number-state $|n_j\rangle$, can result in a transition either to $|\chi; n_j - 1\rangle$ or to $|\chi; n_j + 1\rangle$. In the first instance, the atom absorbs a single photon from mode $j$ and transitions to the higher-energy state $|\chi\rangle$, where the conservation of energy demands that $\mathcal{E}_\chi - \mathcal{E}_\psi \cong \hbar\omega_j$. In the second case, the excited atom moves to the lower-energy state $|\chi\rangle$ by releasing a single photon into the radiation mode $j$, where energy conservation dictates that $\mathcal{E}_\psi - \mathcal{E}_\chi \cong \hbar\omega_j$. If mode $j$ happens to be empty at first, the system could go from $|\psi; 0\rangle$ to $|\chi; 1\rangle$ in a process commonly referred to as spontaneous emission; otherwise, transition from $|\psi; n_j\rangle$ to $|\chi; n_j + 1\rangle$ will be an instance of stimulated emission. While in the case of single-photon absorption from the number-state $|n_j\rangle$ the relevant matrix element of the interaction Hamiltonian $(e/m)\hat{\boldsymbol{p}}\cdot\hat{\boldsymbol{A}}$ is given by $(e/m)\sqrt{\hbar n_j/(2\varepsilon_0 V\omega_j)}\langle\chi|\hat{\boldsymbol{p}}\cdot\hat{\boldsymbol{e}}_j|\psi\rangle$, the corresponding matrix element in the case of spontaneous or stimulated emission will be $(e/m)\sqrt{\hbar(n_j+1)/(2\varepsilon_0 V\omega_j)}\langle\chi|\hat{\boldsymbol{p}}\cdot\hat{\boldsymbol{e}}_j^*|\psi\rangle$. These findings indicate that, in both cases of absorption and stimulated emission, the strength of light-atom interaction (i.e., the squared modulus of the matrix element) should be proportional to the intensity of the incident radiation.

**6.1. Absorption**. In Ref.[9], Sec.1.2.6, it is shown how a constant perturbation, turned on at $t = 0$ with the off-diagonal matrix element $w$, takes an initial state of energy $E_1$ to a different state of energy $E_2$ upon passage of a time $T = \pi/\Omega$, where $\Omega = \sqrt{4|w|^2 + (E_2 - E_1)^2}/\hbar$ is the Rabi frequency.[§] The transition probability is found to peak at $P_{\max} = |w|^2/[|w|^2 + \tfrac{1}{4}(E_2 - E_1)^2]$. In the case of an atom in the state $|\psi\rangle$ moving to a state $|\chi\rangle$ of opposite parity by absorbing a photon from the radiation state $|n_j\rangle$ in mode $j$, we have

$$E_1 = \mathcal{E}_\psi + n_j\hbar\omega_j, \tag{12}$$

$$E_2 = \mathcal{E}_\chi + (n_j - 1)\hbar\omega_j. \tag{13}$$

The absorption probability is thus seen to be maximal when $\mathcal{E}_\chi = \mathcal{E}_\psi + \hbar\omega_j$. Figure 2(a) is a diagrammatic depiction of the various energy levels of the atom during its transition from the ground state $|\psi\rangle$ to the excited state $|\chi\rangle$ via single-photon absorption. It is worth noting the similarities and differences between the present (fully quantum) treatment and the quasi-classical analysis reported, for instance, in Ref.[8], Chapter 2, where a *classical* monochromatic EM field is assumed to excite the atom out of its ground state. In our fully quantum treatment here, once the excitation is turned on at $t = 0$, the matrix element $w$ will be time-independent, and the passage of the joint atom + radiation-state from $|\psi, n_j\rangle$ to $|\chi, n_j - 1\rangle$ becomes a transition between two states of nearly equal energies (i.e., $E_1 \cong E_2$). The proportionality of the matrix element $w$ to $\sqrt{n_j}$, namely,

$$w = (e/m)\sqrt{\hbar n_j/(2\varepsilon_0 V \omega_j)}\,\langle\chi|\hat{\boldsymbol{p}} \cdot \hat{\boldsymbol{e}}_j|\psi\rangle, \tag{14}$$

indicates that the time $T = \pi/\Omega$ at which the absorption probability reaches peak value declines rapidly with an increasing number $n_j$ of incident photons.

---

[§] The symbol $\Omega$ used for the Rabi frequency should not be confused with the same symbol used to specify the differential element $\mathrm{d}\Omega$ of the solid angle in the vicinity of a particular direction, say, $(\theta, \varphi)$, in space, into which photons are emitted or scattered.

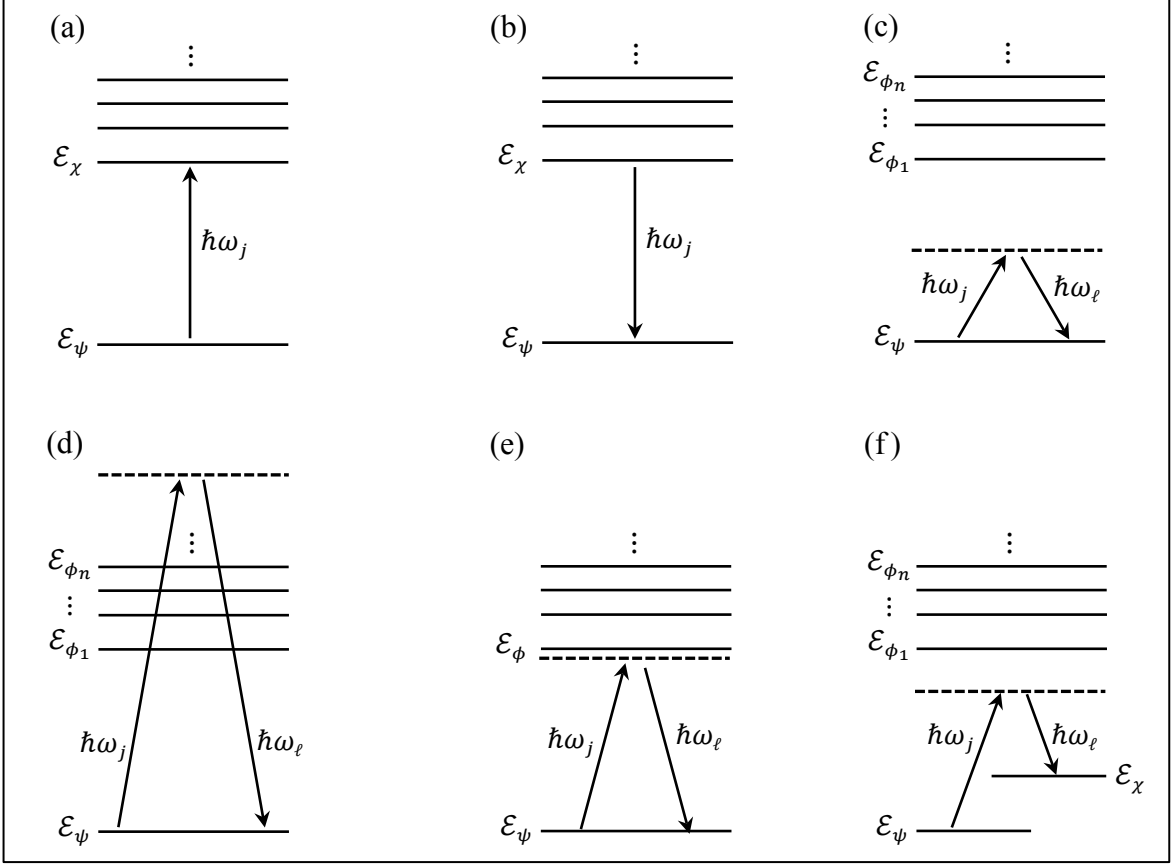

**Fig. 2**. (a) Upon absorbing a single photon of energy $\hbar\omega_j$ from the radiation mode $j$, the atom's electronic state $|\psi\rangle$ transitions to an allowed higher-energy state $|\chi\rangle$, where $\mathcal{E}_\chi - \mathcal{E}_\psi \cong \hbar\omega_j$. (b) Both spontaneous and stimulated emission of a single photon involve transitions from a higher-energy electronic state $|\chi\rangle$ to a lower-energy state $|\psi\rangle$. While in spontaneous emission the photon emerges into an initially empty mode $j$, in the case of stimulated emission the mode $j$ into which the photon is released is initially occupied. Conservation of energy dictates that $\hbar\omega_j \cong \mathcal{E}_\chi - \mathcal{E}_\psi$. (c) In Rayleigh scattering, the incident photon energy $\hbar\omega_j$ is well below the energy gap that separates the ground state $|\psi\rangle$ from all accessible higher-energy states $|\phi\rangle$. The atom returns to its ground state while the scattered photon emerges in a different radiation mode, say, $\ell$, albeit with the same energy, i.e., $\hbar\omega_\ell \cong \hbar\omega_j$. (d) In Thomson scattering, the incident photon energy $\hbar\omega_j$ exceeds the ionization energy of the atom. The atom eventually returns to its ground state, and the scattered photon emerges in a different radiation mode $\ell$ without gaining or losing energy; that is, $\hbar\omega_\ell \cong \hbar\omega_j$. (e) Resonant scattering involves the absorption of an incident photon from mode $j$, followed by its immediate release (via spontaneous emission) into a different radiation mode $\ell$. In the process, the electronic state of the atom first moves from $|\psi\rangle$ to $|\phi\rangle$, an allowed transition with $\mathcal{E}_\phi - \mathcal{E}_\psi \cong \hbar\omega_j$, then returns to $|\psi\rangle$ upon releasing the absorbed photon into mode $\ell$, with $\hbar\omega_\ell = \hbar\omega_j$. (f) Raman scattering is an inelastic scattering process in which virtual transitions to intermediate states $|\phi\rangle$ are followed by a final transition to a state $|\chi\rangle$, where $\mathcal{E}_\chi \neq \mathcal{E}_\psi$. The process entails the capture of an incident photon from mode $j$ and release of a different photon into mode $\ell$, with $\mathcal{E}_\psi + \hbar\omega_j \cong \mathcal{E}_\chi + \hbar\omega_\ell$.

By definition, the intensity of incident radiation is the EM energy crossing a unit area in unit time, namely, the energy content of a volume $c$ of free space. Thus, the intensity of the number-state $|n_j\rangle$, which occupies a spatial volume $V$, turns out to be $(n_j c/V)\hbar\omega_j$. We conclude that the strength $|w|^2$ of the light-atom interaction that appears in the preceding analysis is in direct proportion to the intensity of the incident radiation at the resonant (or nearly resonant) frequency $\omega_j \cong (\mathcal{E}_\chi - \mathcal{E}_\psi)/\hbar$.

**6.2. Stimulated emission**. With reference to Fig.2(b), stimulated emission occurs when, upon interacting with an incident EM field, an atom in the excited state $|\chi\rangle$ moves to a lower-energy state $|\psi\rangle$ by releasing a photon into the same radiation mode as the incident field. Thus, if the incident radiation is the propagating mode $(\omega, \boldsymbol{k}, \hat{\boldsymbol{e}})_j$ in the number-state $|n_j\rangle$, the joint atom-photon state $|\chi, n_j\rangle$ transitions to $|\psi, n_j + 1\rangle$, where the conservation of energy demands that $\mathcal{E}_\chi - \mathcal{E}_\psi \cong \hbar\omega_j$. The interaction Hamiltonian is now

$$(e/m)\sqrt{\hbar/(2\varepsilon_0 V \omega_j)}\,(\hat{\boldsymbol{p}} \cdot \hat{\boldsymbol{e}}_j^*)\hat{a}_j^\dagger, \tag{15}$$

whose matrix element, namely,

$$w = (e/m)\sqrt{\hbar(n_j + 1)/(2\varepsilon_0 V \omega_j)}\,\langle\psi|\hat{\boldsymbol{p}} \cdot \hat{\boldsymbol{e}}_j^*|\chi\rangle, \tag{16}$$

appears in the expression of the Rabi frequency $\Omega = \sqrt{4|w|^2 + (\mathcal{E}_\chi - \mathcal{E}_\psi - \hbar\omega_j)^2}/\hbar$. In accordance with the arguments advanced in Ref.[9], Sec.1.2.6, the maximum transition probability,

$$P_{\max} = \frac{|w|^2}{|w|^2 + \tfrac{1}{4}(\mathcal{E}_\chi - \mathcal{E}_\psi - \hbar\omega_j)^2}, \tag{17}$$

occurs after a time $T = \pi/\Omega$ following the start of the atom-photon interaction. Note the striking similarities between the processes of absorption and stimulated emission and, in particular, the rapid decline of the transition time $T$ with an increasing intensity of the incident radiation at and around the resonance frequency $\omega_j = (\mathcal{E}_\chi - \mathcal{E}_\psi)/\hbar$. The incident intensity $(n_j + 1)\hbar\omega_j c/V$ (i.e., EM energy flux per unit area per unit time) now includes a contribution by vacuum fluctuations in addition to that of the $n_j$ photons that initially resided in mode $j$.

**6.3. Spontaneous emission**. In many respects, spontaneous emission is similar to stimulated emission, albeit with two important differences. First, the stimulating radiation in the case of spontaneous emission is vacuum fluctuations; that is, the stimulating mode $(\omega, \boldsymbol{k}, \hat{\boldsymbol{e}})_j$ is in the number-state $|n_j = 0\rangle$. This means that the joint atom-photon transition takes place between an initially excited state $|\chi, n_j = 0\rangle$ and a final state $|\psi, n_j = 1\rangle$ in which the atom is de-excited by releasing a single photon into the radiation mode $j$ of the surrounding environment; see Fig.2(b). Second, the vacuum mode $j$ is not necessarily unique and, in fact, unless the atom is placed in an environment where the vacuum modes are deliberately curtailed and constrained, the available modes $j$ of the EM field belong to a densely-populated quasi-continuum.

The transition rate $\Gamma$ from a discrete state such as $|\chi, n_j = 0\rangle$ to the quasi-continuum of states surrounding another state, say, $|\psi, n_j = 1\rangle$, under the influence of a constant perturbation with matrix elements $w$ is given by Fermi's golden rule,[8,9] namely,

$$\Gamma = 2\pi|w|^2 \rho_s(\theta, \varphi; \mathcal{E})/\hbar. \tag{18}$$

In the case of spontaneous emission into the (unconstrained) vacuum modes $(\omega, \boldsymbol{k}, \hat{\boldsymbol{e}})_j$, we have

$$\mathcal{E} = \mathcal{E}_\chi - \mathcal{E}_\psi = \hbar\omega_0 \cong \hbar\omega_j. \tag{19}$$

The density of available radiation states (per unit energy per unit solid angle) in the vicinity of a particular emission direction $\hat{\boldsymbol{\kappa}}_j = c\boldsymbol{k}_j/\omega_j$ is given by Eq.(6); here $\hat{\boldsymbol{\kappa}}_j$ is specified by its polar and azimuthal angles $(\theta, \varphi)$, and the spontaneously emitted radiation is restricted to one or the other of a pair of mutually orthogonal polarizations ($\hat{\boldsymbol{e}}_{j,1}$ or $\hat{\boldsymbol{e}}_{j,2}$). For the polarization $\hat{\boldsymbol{e}}_j$ under consideration, the transition matrix element is given by

$$w = (e/m)\sqrt{\hbar/(2\varepsilon_0 V\omega_j)}\,\langle\psi|\hat{\boldsymbol{p}}\cdot\hat{\boldsymbol{e}}_j^*|\chi\rangle. \tag{20}$$

All in all, we find the spontaneous emission rate per unit time per unit solid angle in the chosen direction $(\theta, \varphi)$ with the chosen polarization $\hat{\boldsymbol{e}}_j$ to be

$$\frac{d\varGamma_{\text{sp}}}{d\Omega} = \left(\frac{\omega_0 e^2}{8\pi^2\varepsilon_0\hbar m^2 c^3}\right)|\langle\psi|\hat{\boldsymbol{p}}\cdot\hat{\boldsymbol{e}}_j^*|\chi\rangle|^2. \quad \leftarrow \boxed{\text{subscript sp of } \varGamma \text{ stands for spontaneous}} \tag{21}$$

The above equation reveals that the spontaneous emission rate of an excited atom into its surrounding vacuum varies with the emission direction $(\theta, \varphi)$ in a way that depends on the wavefunctions $\chi(\boldsymbol{r})$ and $\psi(\boldsymbol{r})$ of the initial and final electronic states in conjunction with the polarization $\hat{\boldsymbol{e}}$ of the emitted radiation. For a given pair of electronic states with opposite parities, it is often possible to identify the orthogonal polarizations $\hat{\boldsymbol{e}}_{j,1}$ and $\hat{\boldsymbol{e}}_{j,2}$ such that $\langle\psi|\hat{\boldsymbol{p}}\cdot\hat{\boldsymbol{e}}_{j,1}^*|\chi\rangle \neq 0$ and $\langle\psi|\hat{\boldsymbol{p}}\cdot\hat{\boldsymbol{e}}_{j,2}^*|\chi\rangle = 0$. Needless to say, the overall rate of spontaneous emission is obtained by integrating the differential rate given by Eq.(21) over the entire range of the polar and azimuthal angles $(\theta, \varphi)$ using the identity $d\Omega = \sin\theta\, d\theta d\varphi$.

In Ref.[9], Sec.1.3.2, it is shown that the accessible energies of the quasi-continuum will have a Lorentzian probability distribution of width $\hbar\varGamma$, centered at the most likely energy that the system acquires upon completion of the transition. Considering that the most likely energy of the spontaneously emitted photon is $\hbar\omega_0$ and that the corresponding emission rate is $\varGamma_{\text{sp}}$, the (normalized) lineshape of spontaneously emitted photons will be

$$f_{\text{sp}}(\omega) = \frac{\varGamma_{\text{sp}}/2\pi}{(\omega-\omega_0)^2 + (\varGamma_{\text{sp}}/2)^2}. \tag{22}$$

The full-width-at-half-maximum (FWHM) of this Lorentzian lineshape is the *natural width* of the radiative transition between the electronic states $|\chi\rangle$ and $|\psi\rangle$ of the atom. Thus, with the caveat that the lower energy state $|\psi\rangle$ of the transition is stable, the reciprocal of the lifetime $\varGamma_{\text{sp}}^{-1}$ of the excited state $|\chi\rangle$ coincides with the natural linewidth of the transition.

**7. Elastic and inelastic scattering of photons by atoms in accordance with second-order perturbation theory**. Moving on to second-order perturbations for the $(e/m)\hat{\boldsymbol{p}}\cdot\hat{\boldsymbol{A}}$ interaction Hamiltonian, the system could transition from an initial state $|\psi; n_j, n_\ell\rangle$ to an intermediate state $|\phi; n_j - 1, n_\ell\rangle$ by absorbing a photon from $|n_j\rangle$, then move to a final state $|\chi; n_j - 1, n_\ell + 1\rangle$ by releasing a photon into $|n_\ell\rangle$. The end result of this second-order process is the scattering of one photon from mode $j$ into mode $\ell$. The scattering will be elastic if $\omega_j \cong \omega_\ell$, in which case conservation of energy requires that $\mathcal{E}_\chi = \mathcal{E}_\psi$. The scattering will be inelastic if $\omega_j \neq \omega_\ell$, which requires that $\mathcal{E}_\chi - \mathcal{E}_\psi \cong \hbar(\omega_j - \omega_\ell)$. Since energy conservation need not apply to intermediate states, few restrictions are imposed on such states. In fact, it is permissible for the intermediate state to be $|\phi; n_j, n_\ell + 1\rangle$, which indicates that the virtual transition from the initial atomic state $|\psi\rangle$ to the intermediate state $|\phi\rangle$ is not facilitated by the absorption of a photon from mode $j$. (Recall that the vector potential operator $\hat{\boldsymbol{A}}$ consists of a sum over the various annihilation and creation operators $\hat{a}_j, \hat{a}_\ell, \cdots$ and $\hat{a}_j^\dagger, \hat{a}_\ell^\dagger, \cdots$. Consequently, each instance of the action of $\hat{\boldsymbol{p}}\cdot\hat{\boldsymbol{A}}$ over a joint atom-photon state must result in either creation or annihilation of a single photon.)

In Ref.[9], Sec.1.2.5, it is shown that the effective matrix element for a second-order transition from an initial state $|a\rangle$ to a final state $|b\rangle$ via intermediate states $|c\rangle$ is given by $w_{\text{eff}} = \sum_{c \neq a,b} w_{bc} w_{ca}/(E_c - E_a)$. Thus, the effective matrix element for a second-order transition from $|\psi; n_j, n_\ell\rangle$ to $|\chi; n_j - 1, n_\ell + 1\rangle$ via all accessible electronic states $|\phi\rangle$ is

$$w_{\text{eff}} = \frac{e^2 \hbar}{2\varepsilon_0 m^2 V} \left[\frac{n_j(n_\ell+1)}{\omega_j \omega_\ell}\right]^{1/2} \sum_{\phi \neq \psi, \chi} \left(\frac{\langle \chi|\hat{\mathbf{p}} \cdot \hat{\mathbf{e}}_\ell^*|\phi\rangle \langle \phi|\hat{\mathbf{p}} \cdot \hat{\mathbf{e}}_j|\psi\rangle}{\mathcal{E}_\phi - \mathcal{E}_\psi - \hbar\omega_j} + \frac{\langle \chi|\hat{\mathbf{p}} \cdot \hat{\mathbf{e}}_j|\phi\rangle \langle \phi|\hat{\mathbf{p}} \cdot \hat{\mathbf{e}}_\ell^*|\psi\rangle}{\mathcal{E}_\phi - \mathcal{E}_\psi + \hbar\omega_\ell}\right). \quad (23)$$

In what follows, we shall examine the various scattering regimes that are governed by the second-order effective matrix element of Eq.(23) in addition to the first-order matrix element given by Eq.(8). It will be seen that the value of the incident photon energy $\hbar\omega_j$ relative to the energy gaps $\mathcal{E}_\phi - \mathcal{E}_\psi$ between various atomic states is a decisive factor in determining the scattering cross-section of an atom as well as the linewidth of the scattered photons.

**7.1. Rayleigh scattering.** In Rayleigh scattering, the incident photon energy $\hbar\omega_j$ is well below $\mathcal{E}_\phi - \mathcal{E}_\psi$ for all the electronic states $|\phi\rangle$ that can be reached from the ground state $|\psi\rangle$; see Fig.2(c). The scattering is then an elastic process in which the final atomic state $|\chi\rangle$ coincides with the initial state $|\psi\rangle$, causing the incident and scattered photons to have essentially the same energy; that is, $\omega_\ell \cong \omega_j$. The relevant scattering mechanisms in this case are the second-order process whose perturbing matrix element is given by Eq.(23), and also the first-order direct scattering via the $(e^2/2m)\hat{\mathbf{A}} \cdot \hat{\mathbf{A}}$ Hamiltonian, whose matrix element appears in Eq.(8). When computing the cross-section for Rayleigh scattering, one must add up these matrix elements; the transition rate for scattering into an empty mode (i.e., $n_\ell = 0$) per unit solid angle along a given direction $(\theta, \varphi)$ will then be proportional to the squared modulus of the overall matrix element times the corresponding density of states $\rho_s(\theta, \varphi; \mathcal{E} = \hbar\omega_j)$. When photons of frequency $\omega$ scatter off a two-level atom whose Bohr frequency is $\omega_0$ (i.e., $\mathcal{E}_\phi - \mathcal{E}_\psi = \hbar\omega_0$), the total cross-section for Rayleigh scattering can be shown to be[7]

$$\sigma = (8\pi/3)(2r_e)^2 (\omega/\omega_0)^4, \quad (24)$$

where $2r_e \cong 2.8$ fm is the classical electron diameter. Note that the Rayleigh scattering cross-section varies as the fourth power of the incident frequency $\omega$, implying that short-wavelength radiation is scattered much more strongly than long-wavelength radiation. Hence, in the visible range, scattering is stronger for blue light than for red light, which explains the blue color of the sky and the red of the sunset. The scattering of the sunlight by molecules in the atmosphere is largely Rayleigh scattering associated with electronic molecular resonances in the ultraviolet.[9]

**7.2. Thomson scattering.** In the case of Thomson scattering, the incident photon energy $\hbar\omega_j$ is well above $\mathcal{E}_\phi - \mathcal{E}_\psi$ for all electronic states $|\phi\rangle$ that are accessible from the ground state $|\psi\rangle$. Stated differently, $\hbar\omega_j$ greatly exceeds the ionization energy of the atom in this type of scattering, as depicted in Fig.2(d). Once again, the scattering is an elastic process in which $|\chi\rangle = |\psi\rangle$ and $\omega_\ell \cong \omega_j$. The main difference with the Rayleigh scattering is that the effective matrix element $w_{\text{eff}}$ of Eq.(23) is now fairly small compared to the first-order matrix element $w$ of Eq.(8). This is because the terms $(\mathcal{E}_\phi - \mathcal{E}_\psi + \hbar\omega_\ell)$ and $(\mathcal{E}_\phi - \mathcal{E}_\psi - \hbar\omega_j)$ appearing in the denominators in Eq.(23) can be approximated with $\pm\hbar\omega_j$, respectively, in which case the effective matrix element reduces to

$$w_{\text{eff}} \cong \frac{e^2 [n_j(n_\ell+1)]^{1/2}}{2\varepsilon_0 m^2 V \omega_j^2} \left[\langle \chi|(\hat{\mathbf{p}} \cdot \hat{\mathbf{e}}_j)(\sum_\phi |\phi\rangle\langle\phi|)(\hat{\mathbf{e}}_\ell^* \cdot \hat{\mathbf{p}})|\psi\rangle - \langle \chi|(\hat{\mathbf{p}} \cdot \hat{\mathbf{e}}_\ell^*)(\sum_\phi |\phi\rangle\langle\phi|)(\hat{\mathbf{e}}_j \cdot \hat{\mathbf{p}})|\psi\rangle\right]. \quad (25)$$

Note that $\langle\phi|\hat{\pmb{e}}_j\cdot\hat{\pmb{p}}|\psi\rangle$ is equivalent to $\langle\phi|\hat{\pmb{p}}\cdot\hat{\pmb{e}}_j|\psi\rangle$; that is, $\hat{\pmb{p}}$ can be taken to operate on $|\psi\rangle$ to its right, or on $\langle\phi|$ to its left, without affecting the value of the matrix element under consideration. Similarly, $\langle\phi|\hat{\pmb{e}}_\ell^*\cdot\hat{\pmb{p}}|\psi\rangle = \langle\phi|\hat{\pmb{p}}\cdot\hat{\pmb{e}}_\ell^*|\psi\rangle$. Given that $\sum_\phi|\phi\rangle\langle\phi|$ appearing in Eq.(25) is the identity operator, removing it would further simplify the equation, as follows:

$$w_{\text{eff}} \cong \frac{e^2[n_j(n_\ell+1)]^{1/2}}{2\varepsilon_0 m^2 V \omega_j^2}\big[\langle\chi|(\hat{\pmb{p}}\cdot\hat{\pmb{e}}_j)(\hat{\pmb{e}}_\ell^*\cdot\hat{\pmb{p}}) - (\hat{\pmb{p}}\cdot\hat{\pmb{e}}_\ell^*)(\hat{\pmb{e}}_j\cdot\hat{\pmb{p}})|\psi\rangle\big] = 0. \qquad (26)$$

It is thus seen that the contributions of the $(e/m)\hat{\pmb{p}}\cdot\widehat{\pmb{A}}$ interaction Hamiltonian to Thomson scattering (in accordance with second-order perturbation theory) are negligible, and that the analysis of Sec.5 based on the $(e^2/2m)\widehat{\pmb{A}}\cdot\widehat{\pmb{A}}$ interaction Hamiltonian is quite adequate for describing this type of elastic scattering.

**7.3. Resonant scattering.** Suppose now that the incident photon energy $\hbar\omega_j$ is close to the energy gap $\mathcal{E}_\phi - \mathcal{E}_\psi$ between the ground state $|\psi\rangle$ and one of the allowed states $|\phi\rangle$ of the atom. The small denominator $(\mathcal{E}_\phi - \mathcal{E}_\psi - \hbar\omega_j)$ of the corresponding term in Eq.(23) then ensures the dominance of this term over all the other terms. The scattering mechanism in this case is nothing more nor less than a single photon absorption from mode $j$, followed by spontaneous emission into mode $\ell$, provided that the latter mode is initially empty; that is, $n_\ell = 0$. If, upon absorption and subsequent emission of a photon, the atom returns to its initial state $|\psi\rangle$, we will have $\omega_\ell \cong \omega_j$, and the elastic scattering process is referred to as *resonant scattering*; see Fig.2(e).

**Digression**: It is possible for the atom's final state $|\chi\rangle$ to have a different energy than its initial state $|\psi\rangle$, whence $\hbar\omega_\ell = \hbar\omega_j - (\mathcal{E}_\chi - \mathcal{E}_\psi)$; the resulting inelastic process is then referred to as *Raman scattering*; see Fig.2(f). Raman scattering is particularly useful in molecular physics, where it is used to measure the energy intervals between vibrational or rotational levels belonging to the same electronic ground state. For such levels, a direct $|\psi\rangle \to |\chi\rangle$ transition via photon absorption (or emission) is often forbidden, while Raman scattering is allowed.[9]

We will show in Sec.10 that the scattering cross-section of an atom at exact resonance is $\sigma = \lambda_0^2/2\pi$, where $\lambda_0 = 2\pi c/\omega_0$ is the resonance wavelength. Note that the scattering cross-section at exact resonance depends only on the resonance frequency $\omega_0 = (\mathcal{E}_\phi - \mathcal{E}_\psi)/\hbar$ and not on the specific characteristics of the atomic levels — although the linewidth $\Gamma_{\text{sp}}$ does depend on those characteristics. At visible light frequencies, a typical resonant cross-section is about 16 orders of magnitude greater than that for Thomson scattering ($\lambda_0 \sim 10^{-6}$ m versus $2r_e = 2.8\times 10^{-15}$ m), and the Rayleigh scattering cross-section is smaller than Thomson's by a factor of $(\omega/\omega_0)^4$. The scattering cross-section at resonance, corresponding to dimensions of the order of visible light wavelength, is considerably greater than atomic dimensions (Bohr radius $\cong 0.5$ Å). The large value of the resonant cross-section makes it easy to observe resonant scattering on an atomic vapor with the naked eye. It has even been possible to observe light scattered by a single trapped ion illuminated by a resonant laser. This surprising phenomenon is easy to understand when one remembers that a 0.1 mW laser beam carries $\sim 10^{15}$ photons per second. If the beam has a diameter of 1 cm, the ion will scatter some $10^8$ photons/sec, and an observer placed 30 cm away, will pick up about $10^4$ photons/sec, which is detectable by the human eye.[9]

**8. Relating the electric-dipole Hamiltonian $e\hat{\pmb{r}}\cdot\widehat{\pmb{E}}$ to interaction Hamiltonian $(e/m)\hat{\pmb{p}}\cdot\widehat{\pmb{A}}$.** The electric-dipole Hamiltonian $e\hat{\pmb{r}}\cdot\widehat{\pmb{E}}$ is often used in place of the interaction Hamiltonian $(e/m)\hat{\pmb{p}}\cdot\widehat{\pmb{A}}$. These two operators are closely related and their equivalence can be rigorously established. Here,

we rely on a fairly simple argument to show the equality of their matrix elements $\langle\chi|e\hat{\boldsymbol{r}}\cdot\widehat{\boldsymbol{E}}|\psi\rangle$ and $\langle\chi|(e/m)\hat{\boldsymbol{p}}\cdot\widehat{\boldsymbol{A}}|\psi\rangle$ in a transition that takes the atom from $|\psi\rangle$ to $|\chi\rangle$ or vice-versa in the presence of a resonant (or nearly resonant) photon with energy $\hbar\omega_j \cong \mathcal{E}_\chi - \mathcal{E}_\psi$. Consider the commutator of the electron position operator $\hat{\boldsymbol{r}}$ and the Hamiltonian $\widehat{\mathcal{H}}_0$ of a one-electron atom whose nucleus is located at the origin of the $xyz$ coordinate system, namely,

$$[\hat{\boldsymbol{r}}, \widehat{\mathcal{H}}_0] = \left[\hat{x}\hat{\boldsymbol{x}} + \hat{y}\hat{\boldsymbol{y}} + \hat{z}\hat{\boldsymbol{z}}, \left(\frac{\hat{p}_x^2 + \hat{p}_y^2 + \hat{p}_z^2}{2m}\right) - e\Phi(\hat{\boldsymbol{r}}, t)\right] = (\mathrm{i}\hbar/m)\hat{\boldsymbol{p}}. \qquad (27)^{**}$$

To arrive at the above commutation relation, we have invoked the identities $[\hat{x}, \hat{p}_x] = \mathrm{i}\hbar$, $[\hat{x}, \hat{p}_y] = [\hat{x}, \hat{p}_z] = 0$, $[\hat{x}, \hat{x}] = [\hat{x}, \hat{y}] = [\hat{x}, \hat{z}] = 0$, etc., in addition to slightly more elaborate identities such as $[\hat{x}, \hat{p}_x^2] = 2\mathrm{i}\hbar\hat{p}_x$, which can be proven as follows:

$$\hat{x}\hat{p}_x^2 = (\hat{p}_x\hat{x} + \mathrm{i}\hbar)\hat{p}_x = \hat{p}_x(\hat{x}\hat{p}_x) + \mathrm{i}\hbar\hat{p}_x = \hat{p}_x(\hat{p}_x\hat{x} + \mathrm{i}\hbar) + \mathrm{i}\hbar\hat{p}_x = \hat{p}_x^2\hat{x} + 2\mathrm{i}\hbar\hat{p}_x. \qquad (28)$$

Equation (27) now yields

$$(\mathrm{i}\hbar/m)\langle\chi|\hat{\boldsymbol{p}}|\psi\rangle = \langle\chi|[\hat{\boldsymbol{r}}, \widehat{\mathcal{H}}_0]|\psi\rangle = \langle\chi|(\hat{\boldsymbol{r}}\widehat{\mathcal{H}}_0 - \widehat{\mathcal{H}}_0\hat{\boldsymbol{r}})|\psi\rangle = (\mathcal{E}_\psi - \mathcal{E}_\chi)\langle\chi|\hat{\boldsymbol{r}}|\psi\rangle. \qquad (29)$$

Consequently,

$$\langle\chi|\hat{\boldsymbol{p}}|\psi\rangle = \mathrm{i}[m(\mathcal{E}_\chi - \mathcal{E}_\psi)/\hbar]\langle\chi|\hat{\boldsymbol{r}}|\psi\rangle. \qquad (30)$$

Recall now that, in the long-wavelength approximation within the Schrödinger picture, the vector potential operator $\widehat{\boldsymbol{A}}$ and the electric field operator $\widehat{\boldsymbol{E}}$ of the EM field are given by

$$\widehat{\boldsymbol{A}} \cong \sum_j \sqrt{\hbar/(2\varepsilon_0 V \omega_j)}\,(\hat{\boldsymbol{e}}_j \hat{a}_j + \hat{\boldsymbol{e}}_j^* \hat{a}_j^\dagger), \qquad (31)$$

$$\widehat{\boldsymbol{E}} \cong \mathrm{i}\sum_j \sqrt{\hbar\omega_j/(2\varepsilon_0 V)}\,(\hat{\boldsymbol{e}}_j \hat{a}_j - \hat{\boldsymbol{e}}_j^* \hat{a}_j^\dagger). \qquad (32)$$

Thus, the matrix elements of the $(e/m)\hat{\boldsymbol{p}}\cdot\widehat{\boldsymbol{A}}$ operator will have terms such as $\langle\chi|\hat{\boldsymbol{p}}\cdot\hat{\boldsymbol{e}}_j|\psi\rangle$ and $\langle\chi|\hat{\boldsymbol{p}}\cdot\hat{\boldsymbol{e}}_j^*|\psi\rangle$ with coefficients $(e/m)\sqrt{\hbar/(2\varepsilon_0 V \omega_j)}$, while those of the electric-dipole Hamiltonian $e\hat{\boldsymbol{r}}\cdot\widehat{\boldsymbol{E}}$ will have terms such as $\langle\chi|\hat{\boldsymbol{e}}_j\cdot\hat{\boldsymbol{r}}|\psi\rangle$ and $\langle\chi|\hat{\boldsymbol{e}}_j^*\cdot\hat{\boldsymbol{r}}|\psi\rangle$ with coefficients $\pm\mathrm{i}e\sqrt{\hbar\omega_j/(2\varepsilon_0 V)}$. It is now seen, in the light of Eq.(30), that the matrix elements of the two interaction Hamiltonians coincide when $\hbar\omega_j \cong \mathcal{E}_\chi - \mathcal{E}_\psi$. Stated differently, when the $|\psi\rangle$ to $|\chi\rangle$ atomic transition is resonant (or nearly resonant) with the incident or emitted photon, the corresponding matrix elements of the two interaction Hamiltonians will be essentially the same.

**9. Relating the absorption and emission rates to the electric-dipole interaction coefficient**. Returning to our discussion of single-photon absorption in Sec.6.1, we may now express the Rabi frequency $\Omega$ in terms of the electric-dipole interaction Hamiltonian $e\hat{\boldsymbol{r}}\cdot\widehat{\boldsymbol{E}}$, as follows:

$$\Omega = |w|/\hbar = (e/\hbar)\sqrt{n_j\hbar\omega_j/(2\varepsilon_0 V)}\,|\langle\chi|\hat{\boldsymbol{e}}_j\cdot\boldsymbol{r}|\psi\rangle|. \qquad (33)$$

(The corresponding Rabi frequency for stimulated emission is similar, albeit with $n_j$ replaced by $n_j + 1$.) Assuming the incident light comprises a (more or less uniform) band of frequencies in the immediate vicinity of the atomic transition's resonance frequency $\omega_0 = (\mathcal{E}_\chi - \mathcal{E}_\psi)/\hbar$, we examine a discretized version of this spectrum consisting of a large number of frequencies around $\omega_0$ sampled at $\Delta\omega$ intervals. Given the arbitrarily large volume $V$ of the incident radiation modes, we may assume that $\Delta\omega \gtrsim 2|w|/\hbar$. Writing the overall transition probability from $|\psi\rangle$ to $|\chi\rangle$ as a sum over all the probabilities associated with the discrete frequencies within the band, we will have

$$P_{|\psi\rangle \to |\chi\rangle}(t) \cong \frac{4|w|^2}{\Delta\omega} \int_{\substack{\text{frequency} \\ \text{band}}} \frac{\sin^2\left[\sqrt{4|w|^2+\hbar^2(\omega-\omega_0)^2}\,t/2\hbar\right]}{4|w|^2+\hbar^2(\omega-\omega_0)^2} d\omega. \tag{34}$$

---

[**] The carets over $\hat{\boldsymbol{r}}$, $\hat{\boldsymbol{p}}$, and $\widehat{\mathcal{H}}_0$ indicate that these symbols represent operators. The same is true for the Cartesian components $\hat{x}$, $\hat{y}$, $\hat{z}$ of $\hat{\boldsymbol{r}}$, and also for the components $\hat{p}_x$, $\hat{p}_y$, $\hat{p}_z$ of $\hat{\boldsymbol{p}}$. However, the Cartesian unit-vectors $\hat{\boldsymbol{x}}$, $\hat{\boldsymbol{y}}$, $\hat{\boldsymbol{z}}$ are *not* operators, in spite of the fact that each is adorned with its own caret. It is unfortunate that these conventional representations of operators and unit-vectors occasionally collide and cause confusion. Ideally, the context in which these symbols appear should enable the reader to distinguish one from the other.

The above integrand is well approximated at $\omega = \omega_0$ by $(t/2\hbar)^2$, so long as $|w|t \lesssim \hbar$, and by $\sin^2[\tfrac{1}{2}(\omega-\omega_0)t]/[\hbar(\omega-\omega_0)]^2$ at $\omega = \omega_0 \pm m\Delta\omega$ ($m$ being a positive integer). If the band of incident frequencies is not too narrow, the approximate value of the integral in Eq.(34) will be $(t/2\hbar^2)\int_{-\infty}^{\infty}(\sin x/x)^2 dx = \pi t/2\hbar^2$, in which case the overall $|\psi\rangle \to |\chi\rangle$ transition probability approaches $2\pi(|w|/\hbar)^2 t/\Delta\omega$. We may now invoke Eq.(33) to conclude that the transition rate is

$$\Gamma_{|\psi\rangle \to |\chi\rangle} \cong (\pi e^2 |\langle\chi|\hat{\boldsymbol{e}}_j \cdot \boldsymbol{r}|\psi\rangle|^2/\varepsilon_0\hbar^2)(n_j\hbar\omega_j/V\Delta\omega). \tag{35}$$

Here, $(n_j\hbar\omega_j/V\Delta\omega)$ is the incident radiation's energy density (per unit volume per unit angular frequency) in the vicinity of the resonance frequency $\omega_0$; the other term in Eq.(35), when averaged over all atomic orientations, yields the Einstein $B_{12}$ coefficient as $\pi e^2 r_{12}^2/(3\varepsilon_0\hbar^2)$. Our fully quantum-mechanical derivation of $B_{12}$ is thus seen to be in complete accord with the semi-classical derivation; see Ref.[8], Eq.(2.3.20). Derivation of the quantum formula for the Einstein $B_{21}$ coefficient — a coefficient that quantifies the rate of stimulated emission from an excited atom — follows the same line of reasoning as that which led to Eq.(35); the only difference is that, in the final result, $n_j$ is replaced by $n_j + 1$, indicating that the excited atom is stimulated to release its energy into mode $j$ not only by the $n_j$ incident photons that initially reside in mode $j$, but also by vacuum fluctuations that occur in that mode.

The vacuum fluctuations, of course, are entirely responsible for the spontaneous emission of photons from excited atoms, as described in detail in Sec.6.3. In the light of Eq.(30), the spontaneous emission rate (per unit time per unit solid angle, in the chosen direction $(\theta, \varphi)$ and with the chosen polarization $\hat{\boldsymbol{e}}_j$) given by Eq.(21) can be equivalently expressed in terms of the electric-dipole transition element $\langle\chi|e\hat{\boldsymbol{r}}|\psi\rangle$, as follows:

$$\frac{d\Gamma_{\text{sp}}}{d\Omega} = \left(\frac{e^2\omega_0^3}{8\pi^2\varepsilon_0\hbar c^3}\right)|\langle\chi|\hat{\boldsymbol{e}}_j^* \cdot \hat{\boldsymbol{r}}|\psi\rangle|^2. \tag{36}$$

Taking the polarization vector $\hat{\boldsymbol{e}}_j$ to be real and coplanar with the emitted $k$-vector $\boldsymbol{k}_j$ and the electric dipole moment $-e\langle\hat{\boldsymbol{r}}\rangle_{12}$, we write $|\langle\chi|\hat{\boldsymbol{e}}_j^* \cdot \hat{\boldsymbol{r}}|\psi\rangle| = r_{12}\sin\theta$ and proceed to integrate it over the entire sphere of the space spanned by emitted $k$-vectors. Using the differential solid angle element $d\Omega = \sin\theta\, d\theta d\varphi$ and noting that $\int_{\varphi=0}^{2\pi}\int_{\theta=0}^{\pi}\sin^3\theta\, d\theta d\varphi = 8\pi/3$, we find

$$\Gamma_{\text{sp}} = e^2\omega_0^3 r_{12}^2/(3\pi\varepsilon_0\hbar c^3). \tag{37}$$

The spontaneous emission rate thus obtained with the aid of the full machinery of quantum optics is the well-known Einstein $A$ coefficient that was originally proposed by Albert Einstein in his phenomenological theory of radiation absorption as well as spontaneous and stimulated emission by individual atoms and molecules.[8]

**Observation 1**. Spontaneous emission rates increase rapidly with the radiation frequency, as can be seen from Eq.(37). A transition in the ultraviolet is much stronger than one in the infrared. Typical radiative lifetimes are on the order of nanoseconds in the ultraviolet, ten nanoseconds in the visible, and microseconds in the near infrared. In the far infrared and microwave range, spontaneous emission is a relatively weak process.[9]

**Observation 2**. The rate of spontaneous emission increases faster with frequency than that of stimulated emission, simply because the latter process occurs in a specific mode, whereas the former involves the release of photons into a quasi-continuum of modes whose density rises quadratically with the frequency $\omega_0$ of the radiation. For this reason, it is generally easier to obtain lasing action at longer wavelengths.[9]

**Observation 3**. In spectroscopy, a *forbidden line* is one with a vanishing transition matrix element, say, because the associated levels have the same parity. An interesting example involves the level $2S$ ($n = 2, \ell = 0$) of the hydrogen atom, which cannot decay by electric dipole transition to the ground state $1S$ ($n = 1, \ell = 0$) of the same parity. The only other energy level below $2S$ is $2P_{1/2}$, but the corresponding transition has frequency $\omega_0 = 1$ GHz in the microwave range, whose spontaneous emission rate is, therefore, negligible. The $2S$ level is thus *metastable*, meaning that it is stable despite the fact that its energy is not minimal. The $2S$ state does, in fact, decay with a small probability to the $1S$ state by spontaneous emission of two photons; its decay rate can be calculated by means of a second-order perturbative Hamiltonian as shown in Sec.11. The lifetime associated with this higher-order process is of the order of one second.[9]

**10. Computing the (very large) cross-section of an atom in resonant scattering**. In the case of resonant scattering discussed in Sec.7.3, the dominant contribution to $w_{\text{eff}}$ of Eq.(23) comes from a single intermediate state $|\phi\rangle$ whose energy gap with the ground state $|\psi\rangle$ is very nearly the same as the incident photon energy; that is, $\mathcal{E}_\phi - \mathcal{E}_\psi \cong \hbar\omega_j$. Retaining only the dominant term of Eq.(23) and substituting for $\langle\phi|\hat{\boldsymbol{p}}\cdot\hat{\boldsymbol{e}}_j|\psi\rangle$ from Eq.(30), we find

$$w_{\text{eff}} = -\left(\frac{e^2\hbar\sqrt{n_j\omega_j\omega_\ell}}{2\varepsilon_0 V}\right)\frac{\langle\chi|\hat{\boldsymbol{e}}_\ell^*\cdot\hat{\boldsymbol{r}}|\phi\rangle\langle\phi|\hat{\boldsymbol{e}}_j\cdot\hat{\boldsymbol{r}}|\psi\rangle}{\mathcal{E}_\phi-\mathcal{E}_\psi-\hbar\omega_j}. \tag{38}$$

At exact resonance, the denominator of Eq.(38) vanishes, which is a consequence of the fact that higher-order corrections to the transition amplitude $w_{\text{eff}}$ have not been properly accounted for. A rather simple solution to the problem is to change $\mathcal{E}_\phi$ to $\mathcal{E}_\phi - \mathrm{i}(\hbar\Gamma_{\text{sp}}/2)$, where $\Gamma_{\text{sp}}$ is the spontaneous emission rate given by Eq.(37).[7] This is a reasonable remedy, considering that the time evolution of the excited state $|\phi\rangle$ should be governed by $\exp\{-\mathrm{i}[\mathcal{E}_\phi - \mathrm{i}(\hbar\Gamma_{\text{sp}}/2)]t/\hbar\}$ in order to properly represent the emptying of $|\phi\rangle$ (via spontaneous emission) in accordance with $\exp(-\Gamma_{\text{sp}}t)$. The differential rate of resonant scattering may now be written as follows:

$$\frac{\mathrm{d}\Gamma}{\mathrm{d}\Omega} = 2\pi|w_{\text{eff}}|^2\rho_s(\theta,\varphi;\mathcal{E})/\hbar = 2\pi\left(\frac{e^4\hbar n_j\omega_j\omega_\ell}{4\varepsilon_0^2 V^2}\right)\left[\frac{V\mathcal{E}^2}{(2\pi\hbar c)^3}\right]\frac{|\langle\chi|\hat{\boldsymbol{e}}_\ell^*\cdot\hat{\boldsymbol{r}}|\phi\rangle\langle\phi|\hat{\boldsymbol{e}}_j\cdot\hat{\boldsymbol{r}}|\psi\rangle|^2}{(\mathcal{E}_\phi-\mathcal{E}_\psi-\hbar\omega_j)^2+(\hbar\Gamma_{\text{sp}}/2)^2}. \tag{39}$$

At exact resonance, where $\mathcal{E}_\phi - \mathcal{E}_\psi = \hbar\omega_j$, we have $\mathcal{E} = \hbar\omega_\ell = \hbar\omega_j$. If we assume random orientation for the scattering atom relative to $\hat{\boldsymbol{e}}_j$, we will have $|\langle\phi|\hat{\boldsymbol{e}}_j\cdot\hat{\boldsymbol{r}}|\psi\rangle|^2 = r_{12}^2\cos^2\theta$, which averages to $r_{12}^2/3$. The term $|\langle\chi|\hat{\boldsymbol{e}}_\ell^*\cdot\hat{\boldsymbol{r}}|\phi\rangle|^2 = r_{12}^2\sin^2\theta$, corresponding to a linear polarization $\hat{\boldsymbol{e}}_\ell$ that is coplanar with $\boldsymbol{k}_\ell$ and $\boldsymbol{r}_{12}$, yields $(8\pi/3)r_{12}^2$ upon integration over the full $4\pi$ solid angle in the $k$-space. (There will be no emitted photons with perpendicular polarization to the plane of $\boldsymbol{k}_\ell$ and $\boldsymbol{r}_{12}$, since $\hat{\boldsymbol{e}}_\ell\cdot\boldsymbol{r}_{12} = 0$ for such photons.) Recalling that $\Gamma_{\text{sp}}$ is given by Eq.(37) and evaluating

Eq.(39) under the above circumstances, one finally arrives at $\Gamma = (n_j c/V)(\lambda_0^2/2\pi)$, corresponding to a resonant scattering cross-section $\sigma = \lambda_0^2/2\pi$. (In the literature, the expression of this resonant cross-section is sometimes written as $\sigma = 3\lambda_0^2/2\pi$, which indicates that the polarization $\hat{\boldsymbol{e}}_j$ of the incident light is assumed to be fully aligned with $\boldsymbol{r}_{12} = \langle\phi|\hat{\boldsymbol{r}}|\psi\rangle$.)

**11. Multiphoton processes**. In second-order perturbation theory, an atom under the influence of the interaction Hamiltonian $\widehat{H}_1 = (e/m)(\hbar/2\varepsilon_0\omega_j V)^{½}(\hat{\boldsymbol{p}}\cdot\hat{\boldsymbol{e}}_j)\hat{a}_j$, can transition virtually from an initial state $|\psi;n_j\rangle$ to an intermediate state $|\phi;(n-1)_j\rangle$, then move to a final state $|\chi;(n-2)_j\rangle$, where $\mathcal{E}_\chi - \mathcal{E}_\psi \cong 2\hbar\omega_j$. The effective coupling coefficient for this 2-photon absorption process will then be

$$w_{\text{eff}} = (e/m)^2(\hbar/2\varepsilon_0\omega_j V) \frac{\langle\chi;(n-2)_j|(\hat{\boldsymbol{p}}\cdot\hat{\boldsymbol{e}}_j)\hat{a}_j|\phi;(n-1)_j\rangle \langle\phi;(n-1)_j|(\hat{\boldsymbol{p}}\cdot\hat{\boldsymbol{e}}_j)\hat{a}_j|\psi;n_j\rangle}{\mathcal{E}_\phi - \hbar\omega_j - \mathcal{E}_\psi}. \qquad (40)$$

If there exist several intermediate atomic levels $|\phi_1\rangle$, $|\phi_2\rangle$, etc., which are accessible from both $|\psi\rangle$ and $|\chi\rangle$, then their corresponding (effective) coupling coefficients must be added together. This is a generic example of 2-photon absorption, where a single photon does not have enough energy to cover the gap between $|\psi\rangle$ and $|\chi\rangle$ and, moreover, considering that the initial and final atomic states have the same parity, even a single photon with sufficient energy cannot bring about the $|\psi\rangle \to |\chi\rangle$ transition.

The 2-photon process of stimulated emission is similar to that of 2-photon absorption, except that the coupling coefficient in the case of stimulated emission is given by

$$w_{\text{eff}} = (e/m)^2(\hbar/2\varepsilon_0\omega_j V) \frac{\langle\chi;(n+2)_j|(\hat{\boldsymbol{p}}\cdot\hat{\boldsymbol{e}}_j^*)\hat{a}_j^\dagger|\phi;(n+1)_j\rangle \langle\phi;(n+1)_j|(\hat{\boldsymbol{p}}\cdot\hat{\boldsymbol{e}}_j^*)\hat{a}_j^\dagger|\psi;n_j\rangle}{\mathcal{E}_\phi + \hbar\omega_j - \mathcal{E}_\psi}. \qquad (41)$$

It is also possible to have stimulated emission with only one photon released into the mode $j$ of the incident light, with the other photon being spontaneously emitted into a different mode $\ell$. The stimulated photon (going into the incident mode $j$) can be the first one released, in which case the intermediate state will be the same as that in Eq.(41), namely, $|\phi;(n+1)_j,0_\ell\rangle$, but the final state becomes $|\chi;(n+1)_j,1_\ell\rangle$. Alternatively, spontaneous release can take precedence over stimulated emission, in which case the intermediate state will be $|\phi;n_j,1_\ell\rangle$, and the final state $|\chi;(n+1)_j,1_\ell\rangle$. The effective coupling coefficient in the latter case will be

$$w_{\text{eff}} = (e/m)^2\big[\hbar/2\varepsilon_0(\omega_j\omega_\ell)^{½}V\big] \frac{\langle\chi;(n+1)_j,1_\ell|(\hat{\boldsymbol{p}}\cdot\hat{\boldsymbol{e}}_j^*)\hat{a}_j^\dagger|\phi;n_j,1_\ell\rangle\langle\phi;n_j,1_\ell|(\hat{\boldsymbol{p}}\cdot\hat{\boldsymbol{e}}_\ell^*)\hat{a}_\ell^\dagger|\psi;n_j,0_\ell\rangle}{\mathcal{E}_\phi + \hbar\omega_\ell - \mathcal{E}_\psi}. \qquad (42)$$

Another 2-photon process occurs when an initial excited atomic state $|\psi\rangle$, with assistance from an intermediate state $|\phi\rangle$, goes over to the final state $|\chi\rangle$ by spontaneously emitting a first photon into mode $\ell$ and a second one into mode $\ell$. The effective coupling coefficient in this case will be

$$w_{\text{eff}} = (e/m)^2\big[\hbar/2\varepsilon_0(\omega_\ell\omega_\ell)^{½}V\big] \frac{\langle\chi;1_\ell,1_\ell|(\hat{\boldsymbol{p}}\cdot\hat{\boldsymbol{e}}_\ell^*)\hat{a}_\ell^\dagger|\phi;1_\ell,0_\ell\rangle\langle\phi;1_\ell,0_\ell|(\hat{\boldsymbol{p}}\cdot\hat{\boldsymbol{e}}_\ell^*)\hat{a}_\ell^\dagger|\psi;0_\ell,0_\ell\rangle}{\mathcal{E}_\phi + \hbar\omega_\ell - \mathcal{E}_\psi}. \qquad (43)$$

Aside from the mandatory energy conservation, namely, $\mathcal{E}_\psi - \mathcal{E}_\chi \cong \hbar(\omega_\ell + \omega_\ell)$, the emitted photon frequencies $\omega_\ell$ and $\omega_\ell$ are not constrained in any way; the spontaneously emitted photons may have the same or very different frequencies (i.e., hertzian, microwave, infrared, visible, etc.).

**12. Polarization-entangled photon pair emitted in an atomic radiative cascade**. The calcium atom (Ca) has a ground state $|\varphi_0\rangle$, an intermediate level $|\varphi_1\rangle$, and an excited state $|\varphi_2\rangle$. Both $|\varphi_0\rangle$ and $|\varphi_2\rangle$ are $s$ states, which have even parity and, therefore, cannot be reached directly from one another. In contrast, $|\varphi_1\rangle$ is a $p$ state, which, owing to its odd parity, can serve as an intermediary for $|\varphi_0\rangle \leftrightarrow |\varphi_2\rangle$ transitions. While the ground and excited atomic states have zero angular momenta (i.e., $J_0 = J_2 = 0$), the intermediate state has an angular momentum quantum number $J_1 = 1$, with three degenerate sublevels corresponding to $m_z = -1, 0, +1$, as depicted in Fig.3(a).

A spontaneously-emitted photon that brings down the Ca atom from $|\varphi_2\rangle$ to $|\varphi_{1,-1}\rangle$ with a wavelength of $\lambda_1 = 551$ nm (i.e., $\hbar\omega_1 = \mathcal{E}_2 - \mathcal{E}_1$) will be right-circularly polarized ($\sigma^+$), whereas a second (also spontaneously-emitted) photon that takes the atom from $|\varphi_{1,-1}\rangle$ to $|\varphi_0\rangle$ with a wavelength of $\lambda_2 = 423$ nm (i.e., $\hbar\omega_2 = \mathcal{E}_1 - \mathcal{E}_0$) will have left-circular polarization ($\sigma^-$). Now, if the intermediate state happens to be $|\varphi_{1,+1}\rangle$ instead of $|\varphi_{1,-1}\rangle$, the spontaneously-emitted photon pair will have the same frequencies as before, but their spin angular momenta will be flipped, so that the first emitted photon will be $\sigma^-$ and the second one $\sigma^+$. The matrix elements for the cascade emission process involving the atomic-state transitions $|\varphi_2\rangle \to |\varphi_1\rangle$ and $|\varphi_1\rangle \to |\varphi_0\rangle$ are given by

$$w_{2\to 1} = (e/m)\sqrt{\hbar/(2\varepsilon_0\omega_1 V)}\, \langle\varphi_{1,\pm 1}; 1_j, 0|\hat{\boldsymbol{p}}\cdot\hat{\boldsymbol{e}}_j^*\hat{a}_j^\dagger|\varphi_2; 0, 0\rangle, \tag{44a}$$

$$w_{1\to 0} = (e/m)\sqrt{\hbar/(2\varepsilon_0\omega_2 V)}\, \langle\varphi_0; 1_j, 1_k|\hat{\boldsymbol{p}}\cdot\hat{\boldsymbol{e}}_k^*\hat{a}_k^\dagger|\varphi_{1\pm 1}; 1_j, 0\rangle. \tag{44b}$$

In the above equations, the polarizations $\hat{\boldsymbol{e}}_j$ and $\hat{\boldsymbol{e}}_k$ of the emitted photons are right- and left-circular, namely, $\sigma^\pm = (\hat{\boldsymbol{e}}_x \pm i\hat{\boldsymbol{e}}_y)/\sqrt{2}$, with $\hat{\boldsymbol{e}}_j = \sigma^+$ and $\hat{\boldsymbol{e}}_k = \sigma^-$ if the intermediate state is taken to be $|\varphi_{1,-1}\rangle$ and, conversely, $\hat{\boldsymbol{e}}_j = \sigma^-$ and $\hat{\boldsymbol{e}}_k = \sigma^+$ if the intermediate state is $|\varphi_{1,+1}\rangle$.

Creating a pair of polarization-entangled photons in an atomic radiative cascade entails the pumping of a calcium source from the atomic ground state $|\varphi_0\rangle$ into the excited state $|\varphi_2\rangle$. Given that the parity of both states is even, a direct excitation via electric dipole transition is forbidden;

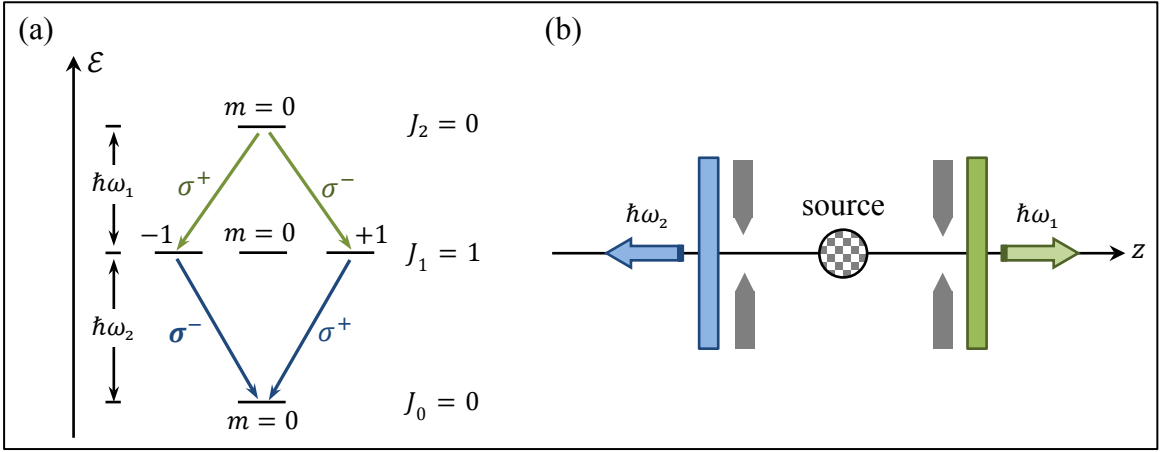

**Fig.3**. (a) The lowest energy levels of the calcium atom are the ground state $|\varphi_0\rangle$, an $s$ state with zero angular momentum ($J_0 = 0$), followed by $|\varphi_1\rangle$, a $p$ state with angular momentum $J_1 = 1$, and $|\varphi_2\rangle$, another $s$ state whose angular momentum is $J_2 = 0$. A two-photon cascade spontaneous emission brings $|\varphi_2\rangle$ to one of the degenerate intermediate states of $|\varphi_1\rangle$, corresponding to $m_z = -1, 0, +1$, and from there to the ground state $|\varphi_0\rangle$. The first transition results in the release of a green photon with energy $\hbar\omega_1$ and either right or left circular polarization (i.e., $\sigma^+$ or $\sigma^-$) depending on whether the intermediate level is $|\varphi_{1,-1}\rangle$ or $|\varphi_{1,+1}\rangle$. The second transition brings about the emission of a violet photon of energy $\hbar\omega_2$ with circular polarization that is opposite to that of the first emitted photon. (b) To produce polarization-entangled photon pairs, a calcium source is brought from the ground state $|\varphi_0\rangle$ to the excited state $|\varphi_2\rangle$ by a two-photon pumping process. The subsequently emitted photons are selected for propagation along the $z$-axis (forward as well as backward directions) by means of apertures placed along their paths. The emitted photons are also selected for their color (i.e., frequency $\omega_1$ or $\omega_2$) via narrowband filters that transmit only the first emitted photon (green) in the $+z$ direction, and the second emitted photon (violet) in the $-z$ direction. In practice, the interference filters are centered on the resonance frequencies $\omega_1$ and $\omega_2$, with passbands that are a little wider than the natural widths of these two lines.

however, a nonlinear two-photon excitation of Ca atoms can be achieved, for example, by means of a krypton ion laser ($\lambda_1' = 406$ nm) and a tunable dye laser ($\lambda_2' = 581$ nm), resulting in a remarkably rapid excitation rate thanks to the nearly resonant nature of this 2-photon absorption process. In a simplified analysis, if one takes the first pump laser to have $n_1$ photons in the $(\omega_1', \hat{e}_1)$ mode, and the second pump laser to have $n_2$ photons in the $(\omega_2', \hat{e}_2)$ mode, then second-order perturbation theory yields the dominant contribution to the effective matrix element for two-photon absorption by calcium atoms, as follows:

$$(e/m)^2 (\hbar/2\varepsilon_0 \omega_1'^{\frac{1}{2}} \omega_2'^{\frac{1}{2}} V) \frac{\langle \varphi_2; n_1-1, n_2-1 | \hat{p} \cdot \hat{e}_2 \hat{a}_2 | \varphi_1; n_1-1, n_2 \rangle \langle \varphi_1; n_1-1, n_2 | \hat{p} \cdot \hat{e}_1 \hat{a}_1 | \varphi_0; n_1, n_2 \rangle}{\mathcal{E}_1 - \hbar\omega_1' - \mathcal{E}_0}. \qquad (45)$$

Needless to say, the small difference between the photon energy $\hbar\omega_1'$ of the krypton ion laser and the gap $\mathcal{E}_1 - \mathcal{E}_0$ separating the atomic levels $|\varphi_0\rangle$ and $|\varphi_1\rangle$ of Ca is a main reason for the high efficiencies achieved by this quasi-resonant two-photon absorption process. Figure 3(b) shows that the spontaneously-emitted cascade photons are selected for their frequencies $\omega_1$ and $\omega_2$ (via narrowband interference filters placed on opposite sides of the source along the z-axis), and also for their propagation directions (by means of apertures positioned along the paths of the emitted radiation). A pair of photons that emerge nearly simultaneously (i.e., within a short time window) in opposite directions will be polarization entangled; that is, they will be in the unfactorizable superposition state $(|\omega_1, \sigma_1^+\rangle |\omega_2, \sigma_2^-\rangle + |\omega_1, \sigma_1^-\rangle |\omega_2, \sigma_2^+\rangle)/\sqrt{2}$.

**13. Atomic energy-level-shifts due to interactions with the vacuum electromagnetic field.** Consider a hydrogen (or hydrogen-like) atom in its ground state $|\varphi_0\rangle$, with its higher energy levels denoted by $|\varphi_1\rangle$, $|\varphi_2\rangle$, etc. The eigenvalues (or energies) of these eigenstates of the unperturbed Hamiltonian $\hat{H}_0 = \hat{p}^2/(2m) - Ze^2/(4\pi\varepsilon_0 \hat{r})$ in the surrounding vacuum are $\mathcal{E}_0$, $\mathcal{E}_1$, $\mathcal{E}_2$, and so on. In first-order perturbation theory, the perturbing Hamiltonian $\hat{H}_1 = (e^2/2m)\hat{A} \cdot \hat{A}$ contains terms such as $(e^2/2m)(\hbar/2\varepsilon_0 \omega_j V)(\hat{e}_j \cdot \hat{e}_j^*)\hat{a}_j \hat{a}_j^\dagger$ that are capable of creation followed by annihilation of a single photon in mode $j$ without changing the atomic state $|\varphi_0\rangle$. All EM vacuum modes $(\omega, k, \hat{e})_j$ are allowed to participate in this process, which starts with the atom-plus-radiation system in the initial state $|\varphi_0, 0\rangle$ and ends up in the same final state $|\varphi_0, 0\rangle$. Given that $\hat{e}_j \cdot \hat{e}_j^* = 1$ for any mode $j$, the corresponding shift in the atomic ground-state energy $\mathcal{E}_0$ in accordance with time-independent perturbation theory[10] is found to be

$$\delta\mathcal{E}_{0,j} = (e^2/2m)(\hbar/2\varepsilon_0 \omega_j V)\langle \varphi_0, 0 | \hat{a}_j \hat{a}_j^\dagger | \varphi_0, 0 \rangle = \hbar e^2/(4\varepsilon_0 m \omega_j V). \qquad (46)$$

The $k$-space consists of spherical shells of radius $k = \omega/c$ and thickness $dk = d\omega/c$, occupied by individual $k$-vectors within each volume element $(2\pi/L)^3$. Thus, the number of EM field modes residing in this shell, taking into account that each $k$-vector is associated with two polarizations $\hat{e}_1$ and $\hat{e}_2$, is $2(L/2\pi)^3 (4\pi k^2) dk = V\omega^2 d\omega/(\pi^2 c^3)$. The energy-level shift $\delta\mathcal{E}_{0,j}$ of the atomic ground state, caused by a single vacuum mode $j$ in conjunction with the perturbation $\hat{H}_1 = (e^2/2m)\hat{A} \cdot \hat{A}$ and given by Eq.(46), must now be multiplied by $V\omega^2 d\omega/(\pi^2 c^3)$, then integrated over $\omega$ (from zero to infinity) to yield the level-shift associated with this particular perturbation. The integral diverges quadratically with $\omega$, calling for the introduction of a cutoff frequency and renormalization techniques, which we shall not pursue in the present paper. We mention in passing that the above contribution to the level shift of the atom is rooted in the kinetic energy gained by the electron due to its immersion in the electric field of the surrounding vacuum.

In second-order perturbation theory, the perturbing Hamiltonian $\hat{H}_1 = (e/m)\hat{p} \cdot \hat{A}$ contains terms such as $(e/m)(\hbar/2\varepsilon_0 \omega_j V)^{\frac{1}{2}}(\hat{p} \cdot \hat{e}_j^*)\hat{a}_j^\dagger$ that account for atomic transitions to different levels, say, $|\varphi_n\rangle$, in addition to creating a single photon in mode $j$ of the EM vacuum. A reverse transition, aided by the operator $(e/m)(\hbar/2\varepsilon_0\omega_j V)^{1/2}(\hat{\boldsymbol{p}}\cdot\hat{\boldsymbol{e}}_j)\hat{a}_j$, then returns the system from $|\varphi_n, 1_j\rangle$ to the ground state $|\varphi_0, 0\rangle$ by annihilating the virtual photon in mode $j$ and returning the atomic state to $|\varphi_0\rangle$. The corresponding change in the energy of the state (in accordance with time-independent perturbation theory[10]) is now found to be

$$\delta\mathcal{E}_{n,j} = (e/m)^2(\hbar/2\varepsilon_0\omega_j V)\frac{|\langle\varphi_n,1_j|(\hat{\boldsymbol{p}}\cdot\hat{\boldsymbol{e}}_j^*)\hat{a}_j^\dagger|\varphi_0,0\rangle|^2}{\mathcal{E}_0 - (\mathcal{E}_n+\hbar\omega_j)}. \tag{47}$$

In contrast to the positive $\delta\mathcal{E}_{0,j}$ of Eq.(46), the level shift $\delta\mathcal{E}_{n,j}$ of Eq.(47) is seen to be negative. It has been suggested that a physical mechanism that contributes to $\delta\mathcal{E}_{n,j}$ is the averaging of the Coulomb potential "seen" by the electron in its vibrational motion under the influence of vacuum fluctuations.[7]

The projections of the overlap vector $\boldsymbol{v} = \langle\varphi_n|\hat{\boldsymbol{p}}|\varphi_0\rangle$ onto the polarization vectors $\hat{\boldsymbol{e}}_1$ and $\hat{\boldsymbol{e}}_2$ associated with each $k$-vector $\boldsymbol{k} = (\omega/c)\hat{\boldsymbol{\kappa}}$, when squared and added together, become equal to $\boldsymbol{v}\cdot\boldsymbol{v}^* - (\boldsymbol{v}\cdot\hat{\boldsymbol{\kappa}})(\boldsymbol{v}^*\cdot\hat{\boldsymbol{\kappa}})$, as shown in the Appendix. Integrating the latter entity over the unit sphere in the $k$-space yields $(8\pi/3)|\langle\varphi_n|\hat{\boldsymbol{p}}|\varphi_0\rangle|^2$. Thus, adding up $\delta\mathcal{E}_{n,j}$ of Eq.(47) over all vacuum modes $j$ entails multiplication by $V\omega^2 d\omega/(\pi^2 c^3)$ of the entity

$$\tfrac{1}{3}(e/m)^2(\hbar/2\varepsilon_0\omega V)|\langle\varphi_n|\hat{\boldsymbol{p}}|\varphi_0\rangle|^2/[\mathcal{E}_0 - (\mathcal{E}_n + \hbar\omega)], \tag{48}$$

followed by integration over $\omega$ from zero to infinity. Once again, the integral diverges (although linearly with $\omega$ in the present case), calling for the introduction of a cutoff frequency and renormalization methods, which are beyond the scope of the present paper.

Needless to say, the atomic energy-level-shift of the ground state $|\varphi_0\rangle$ due to interactions with the vacuum EM field caused by the interaction Hamiltonian $\hat{H}_1 = (e^2/2m)\hat{\boldsymbol{A}}\cdot\hat{\boldsymbol{A}}$ (in first-order perturbation theory) must be added to those produced by the interaction Hamiltonian $\hat{H}_1 = (e/m)\hat{\boldsymbol{p}}\cdot\hat{\boldsymbol{A}}$ (according to second-order perturbation theory and including virtual transitions to all atomic levels $|\varphi_n\rangle$), in order to obtain the lowest-order contribution to the overall level shift.

**14. Concluding remarks**. Invoking elementary methods of quantum electrodynamics, this paper has examined some of the fundamental processes of light-matter interactions. The atom-photon interactions discussed include absorption, stimulated emission, and spontaneous emission, in addition to elastic and inelastic scattering. Section 7 described in some detail the Rayleigh and Thomson scattering of photons from stationary atoms. In Sec.9, we derived the Einstein $A$ and $B$ coefficients using the full machinery of quantum optics. In the case of resonant scattering, where an intermediate state $|\phi\rangle$ of the atom (energy $= \mathcal{E}_\phi$) is accessible from the initial state $|\psi\rangle$ (with energy $\mathcal{E}_\psi$) while the incident photons have energy $\hbar\omega \cong \mathcal{E}_\phi - \mathcal{E}_\psi$, Sec.10 provided a detailed derivation of the (very large) scattering cross-section of the illuminated atom. The scattering is elastic if the atom returns to its initial state $|\psi\rangle$, and inelastic if the energy of the final state $|\chi\rangle$ of the atom is $\mathcal{E}_\chi \neq \mathcal{E}_\psi$.

Nonlinear optical processes such as 2-photon absorption and 2-photon emission are easily analyzed using the same perturbative methods as those used for single-photon processes; this was elaborated in Sec.11. We also explained, in Sec.12, the workings of an atomic radiative cascade in which a pair of polarization-entangled photons are produced. Finally, a brief description of atomic energy-level-shifts in consequence of interactions with vacuum fluctuations was given in Sec.13. All in all, the methods of quantum optics were found to be powerful and versatile, capable of analyzing a broad range of phenomena that are either inaccessible to classical electrodynamics or require resorting to phenomenology (e.g., invoking the mechanical mass-and-spring model of atoms and molecules known as the Lorentz oscillator model), thus obscuring the true underlying physics.

# Appendix

An orthonormal basis in three-dimensional Euclidean space consists of the unit-vector $\hat{\boldsymbol{\kappa}}$, which specifies the propagation direction of an EM plane-wave, and the (generally complex-valued) pair of unit-vectors $\hat{\boldsymbol{e}}_1$ and $\hat{\boldsymbol{e}}_2$, which identify the plane-wave's polarization state. In the Cartesian $xyz$ coordinate system, one may describe the triplet $(\hat{\boldsymbol{\kappa}}, \hat{\boldsymbol{e}}_1, \hat{\boldsymbol{e}}_2)$ in terms of the row vectors $\tilde{\kappa} = (\kappa_x, \kappa_y, \kappa_z)$, $\tilde{e}_1 = (e_{1x}, e_{1y}, e_{1z})$, and $\tilde{e}_2 = (e_{2x}, e_{2y}, e_{2z})$. The orthonormality of the triplet can then be expressed as $\tilde{\kappa}\tilde{\kappa}^T = \tilde{e}_1\tilde{e}_1^{*T} = \tilde{e}_2^T\tilde{e}_2^{*T} = 1$ and $\tilde{\kappa}\tilde{e}_1^{*T} = \tilde{\kappa}\tilde{e}_2^{*T} = \tilde{e}_1\tilde{e}_2^{*T} = 0$. The closure relation in the $(\hat{\boldsymbol{\kappa}}, \hat{\boldsymbol{e}}_1, \hat{\boldsymbol{e}}_2)$ basis is expressed as

$$\tilde{\kappa}^T\tilde{\kappa} + \tilde{e}_1^{*T}\tilde{e}_1 + \tilde{e}_2^{*T}\tilde{e}_2^T = \tilde{I}, \tag{A1}$$

simply because the multiplication of an arbitrary vector $\tilde{v} = \alpha\tilde{\kappa} + \beta\tilde{e}_1 + \gamma\tilde{e}_2$ on the left-hand side of Eq.(A1) results in $\tilde{v} = \tilde{v}$ and, similarly, $\tilde{v}^{*T}$ multiplied on the right-hand side yields $\tilde{v}^{*T} = \tilde{v}^{*T}$. Using the indices $i$ and $j$ to indicate the $x$, or $y$, or $z$ component of a vector, the elements of the $3 \times 3$ matrices in Eq.(A1) are readily seen to satisfy the following identity:

$$\kappa_i \kappa_j + e_{1i} e_{1j}^* + e_{2i} e_{2j}^* = \delta_{ij}. \tag{A2}$$

Here, the Kronecker delta, $\delta_{ij}$, equals 1 for $i = j$, and 0 for $i \neq j$. Suppose now that an arbitrary vector $\boldsymbol{v}$, whose Cartesian components $v_x, v_y, v_z$ are generally complex-valued, is projected onto the unit-vectors $\hat{\boldsymbol{\kappa}}$, $\hat{\boldsymbol{e}}_1$, and $\hat{\boldsymbol{e}}_2$. A useful identity involving the magnitudes of $\boldsymbol{v}$ and its projections $\boldsymbol{v} \cdot \hat{\boldsymbol{\kappa}}$, $\boldsymbol{v} \cdot \hat{\boldsymbol{e}}_1$, and $\boldsymbol{v} \cdot \hat{\boldsymbol{e}}_2$ is proven below.

$$|\boldsymbol{v} \cdot \hat{\boldsymbol{e}}_1|^2 + |\boldsymbol{v} \cdot \hat{\boldsymbol{e}}_2|^2 = \sum_{i=x,y,z}(v_i e_{1i})\sum_{j=x,y,z}(v_j^* e_{1j}^*) + \sum_{i=x,y,z}(v_i e_{2i})\sum_{j=x,y,z}(v_j^* e_{2j}^*)$$

$$= \sum_i \sum_j v_i v_j^*(e_{1i}e_{1j}^* + e_{2i}e_{2j}^*) = \sum_i \sum_j v_i v_j^*(\delta_{ij} - \kappa_i\kappa_j) \quad \leftarrow \text{see Eq.(A2)}$$

$$= \boldsymbol{v} \cdot \boldsymbol{v}^* - (\boldsymbol{v} \cdot \hat{\boldsymbol{\kappa}})(\boldsymbol{v}^* \cdot \hat{\boldsymbol{\kappa}}). \tag{A3}$$

In hindsight, the above identity should not come as a surprise, since it simply asserts that the squared magnitude of $\boldsymbol{v}$ equals the sum of the squared magnitudes of its projections onto the triplet of mutually orthogonal unit-vectors $(\hat{\boldsymbol{\kappa}}, \hat{\boldsymbol{e}}_1, \hat{\boldsymbol{e}}_2)$.

# References

1. M. Born and E. Wolf, *Principles of Optics* (7th expanded edition), Cambridge University Press, Cambridge, United Kingdom (1999).
2. J. D. Jackson, *Classical Electrodynamics* (3rd edition), Wiley, New York (1999).
3. R. P. Feynman, R. B. Leighton, and M. Sands, *The Feynman Lectures on Physics*, Addison-Wesley, Massachusetts (1965).
4. M. Mansuripur, *Field, Force, Energy and Momentum in Classical Electrodynamics* (revised edition), Bentham Science Publishers, Sharjah (2017).
5. M. Mansuripur, "Insights into the behavior of certain optical systems gleaned from Feynman's approach to quantum electrodynamics," *Proceedings of SPIE* **12197**, Plasmonics: Design, Materials, Fabrication, Characterization, and Applications XX, **1219703** (2022).
6. M. Mansuripur, "Absorption and Stimulated Emission by a Thin Slab Obeying the Lorentz Oscillator Model," *Japanese Journal of Applied Physics* **58**, SKKB02 (2019).
7. C. Cohen-Tannoudji, J. Dupont-Roc, and G. Grynberg, *Atom-Photon Interactions: Basic Processes and Applications*, Wiley, New York (1992).
8. R. Loudon, *The Quantum Theory of Light*, 3rd edition, Oxford University Press, Oxford, United Kingdom (2000).
9. G. Grynberg, A. Aspect, and C. Fabre, *Introduction to Quantum Optics*, Cambridge University Press, Cambridge, United Kingdom (2010).
10. D. Bohm, *Quantum Theory*, Prentice-Hall, New Jersey (1951).